\documentclass[a4paper,11pt]{article}
\pdfoutput=1
\usepackage{jheppub}

\usepackage{amssymb,amsmath}
\usepackage[normalem]{ulem}
\usepackage[utf8x]{inputenc}
\usepackage{slashed}
\usepackage{graphicx}
\usepackage{here}
\usepackage{color}
\usepackage{csquotes} 
\usepackage{comment}
\usepackage{mathrsfs}
\usepackage{float}
\usepackage{ascmac}
\usepackage{multirow}
\usepackage{longtable}
\usepackage{bm}
\usepackage{tikz}

\usepackage[italicdiff]{physics}

\usepackage[hang,small,bf]{caption}
\usepackage[subrefformat=parens]{subcaption}
\makeatletter
\newcommand*\rel@kern[1]{\kern#1\dimexpr\macc@kerna}
\newcommand*\widebar[1]{%
  \begingroup
  \def\mathaccent##1##2{%
    \rel@kern{0.8}%
    \overline{\rel@kern{-0.8}\macc@nucleus\rel@kern{0.2}}%
    \rel@kern{-0.2}%
  }%
  \macc@depth\@ne
  \let\math@bgroup\@empty \let\math@egroup\macc@set@skewchar
  \mathsurround\z@ \frozen@everymath{\mathgroup\macc@group\relax}%
  \macc@set@skewchar\relax
  \let\mathaccentV\macc@nested@a
  \macc@nested@a\relax111{#1}%
  \endgroup
}
\makeatother

\numberwithin{equation}{section}

\preprint{
\begin{minipage}{5cm}
\small
\flushright
EPHOU-24-017\\
KYUSHU-HET-304
\end{minipage}}

\title{Quantum aspects of non-invertible flavor symmetries in intersecting/magnetized D-brane models}

\author{Shuta Funakoshi$^{1}$,} 
\author{Tatsuo Kobayashi$^{2}$, and} 
\author{Hajime Otsuka$^{1}$} 
\affiliation{
$^1$Department of Physics, Kyushu University, 744 Motooka, Nishi-ku, Fukuoka 819-0395, Japan}
\affiliation{
$^2$Department of Physics, Hokkaido University, Sapporo 060-0810, Japan}
\emailAdd{funakoshi.shuta@phys.kyushu-u.ac.jp}
\emailAdd{kobayashi@particle.sci.hokudai.ac.jp}
\emailAdd{otsuka.hajime@phys.kyushu-u.ac.jp}

\abstract{
We discuss selection rules of chiral matters in type IIA intersecting and IIB magnetized D-brane models on toroidal orbifolds. 
Since the chiral matters on toroidal orbifolds are labeled by a certain conjugacy class of the gauged orbifold group, the selection rules involve non-trivial fusion rules. 
We find that the representation of the chiral matters is described by a $D_4$ flavor symmetry for an even number of magnetic fluxes or winding numbers at tree level. Furthermore, the $D_4$ symmetry still remains even when we take into account loop effects. We also study non-perturbative effects such as D-brane instantons. 
}

\makeatletter
\gdef\@fpheader{}
\makeatother

\begin{document}

\maketitle

\section{Introduction}

Non-invertible symmetries are of particular interest 
not only for two-dimensional (2D) conformal field theory (CFT) but also for four-dimensional (4D) quantum field theory. 
The product of topological operators $U_i$ defining the non-invertible symmetry obeys a general fusion rule $U_i \otimes U_j = \sum_k c_{ij}^k U_k$ with $c_{ij}^k$ being constants, as known in Verlinde lines of 2D rational CFTs \cite{Verlinde:1988sn,Moore:1988qv,Moore:1989yh} (for more details, see, e.g., Refs.~\cite{Schafer-Nameki:2023jdn,Shao:2023gho}). 

For an ordinary symmetry described by a group $G$, the representation of fields is specified by a certain representation of $G$, but in a more generic case, the fields are labeled by a conjugacy class of $G$. 
It was indeed known that in non-Abelian orbifolds with non-Abelian group $\Gamma$ \cite{Dixon:1986qv}, twisted sectors are labeled by conjugacy classes $[g_i]$ of the gauging of $\Gamma$, and the fusion of two conjugacy classes is given by the sum of all the conjugacy classes, i.e., $[g_i]\otimes [g_j] = \sum_k c_{ij}^k [g_k]$ \cite{Hamidi:1986vh,Dijkgraaf:1989hb}. 
This selection rule constrains the interaction of fields in the Lagrangian at tree level, but it will be violated at loop order and the remaining selection rule is controlled by an invertible symmetry, i.e., the abelianization of $\Gamma$, $\Gamma/[\Gamma, \Gamma]$ \cite{McNamara:2021cuo,Heckman:2024obe,Kaidi:2024wio}. 
Such a selection rule associated with group-like symmetry is of particular interest in flavor physics because the twisted sectors can be identified with chiral matters in the context of 4D low-energy effective action of heterotic string theory on orbifolds and the non-invertible symmetries play a role as flavor symmetries \cite{Kobayashi:2004ya,Kobayashi:2006wq,Beye:2014nxa}. 

In this paper, we study the selection rules of group-like symmetry in the context of type IIA intersecting and type IIB magnetized D-brane models on toroidal orbifolds. 
Since a topological operator on the underlying torus is not a gauge-invariant quantity under the orbifold group $\Gamma$, it is expected that the operators on toroidal orbifolds can be constructed in a $\Gamma$-invariant way. 
In the context of type IIB magnetized D-brane on $T^2/\mathbb{Z}_N$ with $U(1)$ magnetic flux, momentum operators are indeed restricted to a $\mathbb{Z}_N$ invariant form with non-trivial fusion rules \cite{Kobayashi:2024yqq}. 
It turned out that degenerate chiral zero modes induced by background magnetic fluxes transform under this non-invertible symmetry, yielding their selection rules. 
The obtained selection rules on $T^2/\mathbb{Z}_2$ lead to the $D_4$ symmetry of chiral matters when the magnetic flux is even. 
On the other hand, when the magnetic flux is odd, 
there is no invertible symmetry, but we still have non-trivial coupling selection rules. 
The purpose of this paper is twofold. 
First, we revisit the non-invertible symmetry in type IIB magnetized D-brane models from the viewpoint of its T-dual intersecting D-brane models. 
Since the 2D CFT picture is clear in intersecting D-brane models, the flavor symmetry acting on the twisted fields can be understood in a way similar to the analysis of Ref. \cite{Kobayashi:2024yqq}. 
Second, we study quantum corrections to the selection rules of chiral matters. 
By deriving the selection rules at loop level, it is found that the $D_4$ symmetry realized at tree level still remains even when we take into account loop effects. Furthermore, we discuss the fate of flavor symmetry associated with the non-invertible symmetry at the non-perturbative level. Even in that case, a certain  symmetry is still allowed in the 4D low-energy effective action.

This paper is organized as follows. 
In Sec. \ref{sec:generic-loop}, we review the selection rules of group-like symmetry at loop level, following Refs.~\cite{Heckman:2024obe,Kaidi:2024wio}. 
In Sec. \ref{sec:loop}, we revisit the non-invertible symmetry in type IIA intersecting D-brane models after reviewing that in its T-dual magnetized D-brane models. Furthermore, we apply the discussion of Sec. \ref{sec:generic-loop} to the type IIB magnetized D-brane system in Sec. \ref{sec:IIB_loop}. 
In Sec. \ref{sec:np}, we discuss the selection rules at the non-perturbative level by studying the D-brane instanton effects. 
Finally, Sec. \ref{sec:con} is devoted to the conclusions and discussions.  In Appendix \ref{app}, we list an explicit form of 3-point couplings in a concrete magnetized D-brane model.  In Appendix \ref{app:dijk}, we summarize the transformations of 3-point couplings under the non-invertible symmetry in magnetized D-brane models.

\section{Perturbative effects on non-invertible symmetries}
\label{sec:generic-loop}

In this section, we study loop effects on selection rules due to the non-invertible symmetries. 
We follow the discussion in Ref.~\cite{Kaidi:2024wio} with a slight extension.

We start with the conventional symmetry, which corresponds to the group $G$.
Each field $\phi_i$ has a definite representation $g_i$ of $G$.
Their Lagrangian must be invariant, and bare couplings $\phi_1\cdots \phi_N$ 
are allowed only if their representations satisfy the following conditions:
\begin{align}
\label{eq:selec-rule-1}
    g_1\cdots g_N = e,
\end{align}
where $e$ denotes the identity in $G$.
Let us discuss the tree-level diagram including two vertices, which are connected 
by an internal line.
We denote incoming modes and outgoing modes of the internal line by $\phi_I$ and $\phi_I^*$, and they correspond to the representations, $h_I$ and $h_I^{-1}$, respectively.
This tree-level diagram corresponds to the product of representations: 
\begin{align}
    g_1\cdots g_i h_I h_I^{-1}g_{i+1}\cdots g_N = e.
\end{align}
This selection rule is consistent with Eq.~(\ref{eq:selec-rule-1}).
This holds true for all tree-level diagrams with many vertices and loop diagrams 
if there is no anomaly.\footnote{Symmetry breaking by anomaly is rather non-perturbative effects.}
Thus, the selection rules are not violated perturbatively.

Now, let us study gauging the above theory.
We introduce some transformations $x$ of group elements: 
\begin{align}
    g_i \to xg_ix^{-1},
\end{align}
where $x$ may be a single transformation or include several transformations.
Then, we define the class $[g_i]$, which is the set including $g_i$ and all of $xg_ix^{-1}$.
For example, when $x$ corresponds to all of the elements of $G$, 
the class $[g_i]$ is nothing but the conjugacy class, to which $g_i$ belongs.
Another example of $x$ is the outer automorphism of $G$.
After gauging by $x$, each field $\phi_i$ is labeled not by the single representation $g_i$, but by the class $[g_i]$, which includes more than one representation.
In general, the classes $[g_i]$ satisfy non-trivial fusion rules.

Let us examine the coupling selection rules.
We have started with $G$-invariant theory.
The allowed coupling terms in the bare Lagrangian satisfy the rule (\ref{eq:selec-rule-1}).
This selection rule is explained by terminology of $[g_i]$ as follows.

\paragraph{Selection rule:}
The coupling term $\phi_1\cdots \phi_N$ is allowed in the bare Lagrangian only if 
one can choose $g_i$ from $[g_i]$, which satisfy the selection rule (\ref{eq:selec-rule-1}).

\begin{figure}
\begin{minipage}[t]{0.45\textwidth}
\begin{tikzpicture}[thick]

\node[circle, draw, fill=gray!30, minimum size=1.2cm] (A) at (0, 0) {};
\node[circle, draw, fill=gray!30, minimum size=1.2cm] (B) at (3, 0) {};

\draw[blue, thick] (A.east) -- (B.west);
\draw[red, thick] (1.5, 0.15) -- (1.5, -0.15); %

\draw[<-] (A) -- ++(-1, 1) node[left] {$[g_1]$};
\draw[<-] (A) -- ++(-1, 0.5) node[left] {$[g_2]$};
\draw[<-] (A) -- ++(-1, -0.2) node[left] {$\vdots$};
\draw[<-] (A) -- ++(-1, -1) node[left] {$[g_{i}]$};

\draw[<-] (B) -- ++(1, 1) node[right] {$[g_{i+1}]$};
\draw[<-] (B) -- ++(1, 0.5) node[right] {$[g_{i+2}]$};
\draw[<-] (B) -- ++(1, -0.2) node[right] {$\vdots$};
\draw[<-] (B) -- ++(1, -1) node[right] {$[g_N]$};

\end{tikzpicture}
\end{minipage}
\begin{minipage}[t]{0.45\textwidth}
\begin{tikzpicture}[thick]

\node[circle, draw, fill=gray!30, minimum size=1.2cm] (A) at (0, 0) {};
\node[circle, draw, fill=gray!30, minimum size=1.2cm] (B) at (4, 0) {};

\draw[<-] (A) -- ++(-1, 1) node[left] {$[g_1]$};
\draw[<-] (A) -- ++(-1, 0.5) node[left] {$[g_2]$};
\draw[<-] (A) -- ++(-1, -0.2) node[left] {$\vdots$};
\draw[<-] (A) -- ++(-1, -1) node[left] {$[g_{i}]$};
\draw[<-] (A) -- ++(1.75, 0) node[midway, above] {$[h_{I}]$};

\draw[<-] (B) -- ++(1, 1) node[right] {$[g_{i+1}]$};
\draw[<-] (B) -- ++(1, 0.5) node[right] {$[g_{i+2}]$};
\draw[<-] (B) -- ++(1, -0.2) node[right] {$\vdots$};
\draw[<-] (B) -- ++(1, -1) node[right] {$[g_N]$};
\draw[<-] (B) -- ++(-1.75, 0) node[midway, above] {$[h_I]^{-1}$};

\end{tikzpicture}
\end{minipage}
    \caption{A tree-level diagram with the $(k + 1)$-vertices and external legs labeled by $[g_1], \cdots, [g_N]$ in the left figure. By cutting one of the internal lines of $\phi_I\phi_I^\ast$ corresponding to $[h_I][h_I^{-1}]$ (red line in the left figure), we have two tree-level diagrams whose vertex numbers are less than $k+1$ in the right figure.}
    \label{fig:tree}
\end{figure}
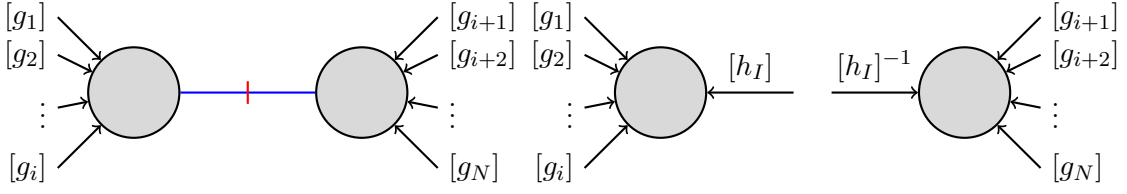

\paragraph{}
When diagrams include internal lines, the selection rule may change.
The important point is as follows.
When the incoming mode $\phi_I$ of the internal line belongs to $[h_I]$, 
the outgoing mode $[\phi_I^*]$ belongs to   $[h_I^{-1}]$.
All of the elements $h_I \in [h_I]$ and $h_I'^{-1} \in [h_I^{-1}]$ do not satisfy 
$h_I h'^{-1}_I=e$.
However, the tree-level diagrams with many internal lines also satisfy the same rule.
We can show this by means of mathematical induction following Ref.~{\cite{Kaidi:2024wio}}.
The single vertex diagrams correspond to coupling terms in the bare Lagrangian.
Obviously, they satisfy the rule (\ref{eq:selec-rule-1}).
Suppose that the tree-level diagrams with $k$ vertices satisfy the 
rule (\ref{eq:selec-rule-1}).
Then, we examine the tree-level diagrams with $k+1$ vertices. 
See Figure \ref{fig:tree}.
We cut one of internal lines of $\phi_I\phi_I^*$, which corresponds to $[h_I][h_I^{-1}]$.
After cutting, we have two tree-level diagrams, whose vertex numbers are less than 
$k+1$.
They must satisfy the rule \eqref{eq:selec-rule-1} by the above assumption.
We find 
\begin{align}
    A_1=g_1\cdots g_i h_I=e, \qquad A_2=h_I'^{-1}g_{i+1} \cdots g_N = e,
\end{align}
with $h_I \in [h_I]$ and $h_I'^{-1} \in [h_I^{-1}]$.
The element $h_I'^{-1}$ may not satisfy $h_Ih_I'^{-1}=e$.
However, we can find the transformation $x$ such that $h_I^{-1}=x h_I'^{-1}x^{-1}$  
and $h_Ih_I^{-1}=e$.
Then, we transform $A_2$ by $x$ as 
\begin{align}
      x A_2 x^{-1}= h_I^{-1}(xg_{i+1}x^{-1}) \cdots (xg_Nx^{-1})=e. 
\end{align}
Note that $xg_ix^{-1} \in [g_i]$.
We multiply $A_1$ and $xA_2x^{-1}$ so as to find 
\begin{align}
    A_1xA_2x^{-1} 
    &= g_1\cdots g_i (xg_{i+1}x^{-1}) \cdots (xg_Nx^{-1}) 
    =e.
\end{align}
Therefore, the tree-level diagrams with $k+1$ vertices also satisfy 
the above selection rule.

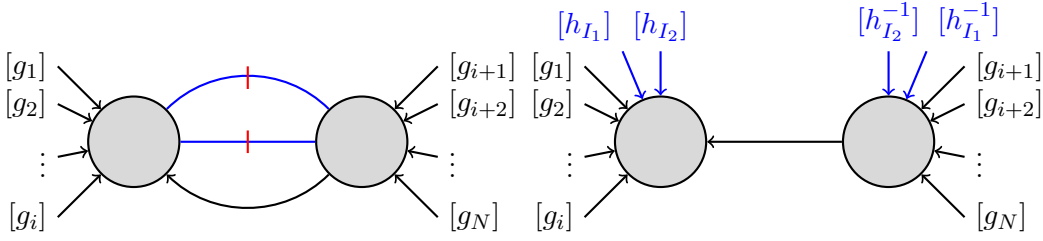
\begin{figure}[H]
\begin{minipage}[t]{0.45\textwidth}
\begin{tikzpicture}[thick]

\node[circle, draw, fill=gray!30, minimum size=1.2cm] (A) at (0, 0) {};
\node[circle, draw, fill=gray!30, minimum size=1.2cm] (B) at (3, 0) {};

\draw[blue, thick] (A.east) -- (B.west);
\draw[red, thick] (1.5, 0.15) -- (1.5, -0.15); %

\draw[<-, thick] (A.south east) to[out=-45, in=-135] node[midway, below] {} (B.south west);

\draw[thick, blue] (A.north east) to[out=45, in=135] node[midway, above] {} (B.north west);
\draw[red, thick] (1.5, 1.0) -- (1.5, 0.7); %

\draw[<-] (A) -- ++(-1, 1) node[left] {$[g_1]$};
\draw[<-] (A) -- ++(-1, 0.5) node[left] {$[g_2]$};
\draw[<-] (A) -- ++(-1, -0.2) node[left] {$\vdots$};
\draw[<-] (A) -- ++(-1, -1) node[left] {$[g_{i}]$};

\draw[<-] (B) -- ++(1, 1) node[right] {$[g_{i+1}]$};
\draw[<-] (B) -- ++(1, 0.5) node[right] {$[g_{i+2}]$};
\draw[<-] (B) -- ++(1, -0.2) node[right] {$\vdots$};
\draw[<-] (B) -- ++(1, -1) node[right] {$[g_N]$};

\end{tikzpicture}
\end{minipage}
\begin{minipage}[t]{0.45\textwidth}
\begin{tikzpicture}[thick]

\node[circle, draw, fill=gray!30, minimum size=1.2cm] (A) at (0, 0) {};
\node[circle, draw, fill=gray!30, minimum size=1.2cm] (B) at (3, 0) {};

 Horizontal line connecting the two circles
\draw[<-] (A) -- (B) node[midway, above] {};

\draw[<-] (A) -- ++(-1, 1) node[left] {$[g_1]$};
\draw[<-] (A) -- ++(-1, 0.5) node[left] {$[g_2]$};
\draw[<-] (A) -- ++(-1, -0.2) node[left] {$\vdots$};
\draw[<-] (A) -- ++(-1, -1) node[left] {$[g_{i}]$};
\draw[<-, blue] (A) -- ++(-0.5, 1.2) node[above left] {$[h_{I_1}]$};
\draw[<-, blue] (A) -- ++(0, 1.2) node[above] {$[h_{I_2}]$};

\draw[<-] (B) -- ++(1, 1) node[right] {$[g_{i+1}]$};
\draw[<-] (B) -- ++(1, 0.5) node[right] {$[g_{i+2}]$};
\draw[<-] (B) -- ++(1, -0.2) node[right] {$\vdots$};
\draw[<-] (B) -- ++(1, -1) node[right] {$[g_N]$};
\draw[<-, blue] (B) -- ++(0.5, 1.2) node[above right] {$[h_{I_1}^{-1}]$};
\draw[<-, blue] (B) -- ++(0, 1.2) node[above] {$[h_{I_2}^{-1}]$};

\end{tikzpicture}
\end{minipage}
    \caption{A two-loop diagram with the $(k + 1)$-vertices and external legs labeled by $[g_1], \cdots, [g_N]$ in the left figure. By cutting two of internal lines corresponding to $[h_{I_1}][h_{I_1}^{-1}]$ and $[h_{I_2}][h_{I_2}^{-1}]$ (red line in the left figure), we have a tree-level diagram with external legs labeled by $[g_1], \cdots, [g_N], [h_{I_1}], [h_{I_2}], [h_{I_1}^{-1}], [h_{I_2}^{-1}]$ in the right figure.}
    \label{fig:loop}
\end{figure}

Next, let us examine loop diagrams.
For example, we focus on the two-loop diagrams shown in Figure \ref{fig:loop}.
We cut two internal lines, which correspond to $[h_{I_1}][h_{I_1}^{-1}]$
and $[h_{I_2}][h_{I_2}^{-1}]$. 
Since it is the tree-level diagram after cutting, 
it should satisfy the selection rule:
\begin{align}
    A_1=g_1\cdots g_i h_{I_1}h_{I_2}=e, \qquad A_2=h_{I_1}'^{-1}h_{I_2}'^{-1}g_{i+1} \cdots g_N = e,
\end{align}
with $h_{I_1,I_2} \in [h_{I_1,I_2}]$ and $h_{I_1,I_2}'^{-1} \in [h_{I_1,I_2}^{-1}]$.
Their multiplication leads to the following equation:
\begin{align}
    g_1\cdots g_ig_{i+1} \cdots g_N = (h_{I_1}^{-1}h_{I_1}')(h_{I_2}^{-1}h_{I_2}').
\end{align}
The right hand side does not always satisfy $(h_{I_1}^{-1}h_{I_1}')(h_{I_2}^{-1}h_{I_2}')=e$, because of ambiguity due to $x$, i.e., $h'_{I_1,I_2}=xh_{I_1,I_2}x^{-1}$. 
It is the important point that $[h_I]$ includes more than one element $h_i$.
If the theory includes $[g_i]$, whose element is only a single element $g_i$, 
i.e., $xg_ix^{-1}=g_i$, 
its selection rule is not violated by loop effects even after gauging.

We have shown that the selection rule due to the non-invertible symmetries can 
be violated by loop effects.
We give a comment in the case that a subgroup symmetry $H$ of $G$ remains after gauging.
Each field $\phi_i$ is labeled by the class $[g_i]$.
However, if the subgroup symmetry $H$ of $G$ remains, 
each class $[g_i]$ has a definite representation of $H$, $r_i$.
Let us denote it by $[g_i]_{r_i}$.
Only for $r_i$, the conventional selection rule of $H$ holds true.
Alternatively, one can say that $[g_i]_{r_i}$ has a single and definite 
quantum number under $H$.
Note that $xr_ix^{-1}=r_i$ and $[g_i]_{xr_ix^{-1}}=[g_i]_{r_i}$.
Their selection rule of $H$ is not violated by loop effects, although 
the other selection rule due to gauging $G$ by $x$ may be violated.

\section{Non-invertible symmetries in D-brane models at perturbative level}
\label{sec:loop}

In this section, we first briefly review the flavor symmetries of matter zero modes in type IIB magnetized D-brane models in Sec. \ref{sec:flavor_IIB}. 
After gauging the orbifold twist, the flavor symmetry on the orbifold is originated from the non-invertible symmetry following Ref. \cite{Kobayashi:2024yqq}. 
In Sec. \ref{sec:flavor_IIA}, we next extend the analysis of Sec. \ref{sec:flavor_IIB} 
to the T-dual type IIA intersecting D-brane models on toroidal orbifolds. 
Finally, we analyze the effects of quantum corrections to the flavor symmetries of matter fields in Sec. \ref{sec:IIB_loop}.

\subsection{Flavor symmetries in magnetized D-brane models}
\label{sec:flavor_IIB}

In this section, we review the non-invertible symmetry in the context of type IIB magnetized D-brane models on $T^2/\mathbb{Z}_2$. 
Specifically, let us start with two stacks of magnetized D-branes $a$ and $b$ on $T^2$ whose complex coordinate is given by $z=y_1 + \tau y_2$, where $y_1$ and $y_2$ are real coordinates and $\tau$ is the complex structure. 
Under the $U(1)$ background magnetic flux:
\begin{align}
    F = \frac{\pi}{{\rm Im}(\tau)} (m_a X_a + m_b X_b) dz \wedge d\Bar{z},
\end{align}
with 
\begin{align}
        X_a =
        \begin{pmatrix}
        1 & 0\\
        0 & 0 \\
    \end{pmatrix}
    ,\qquad
        X_b =
        \begin{pmatrix}
        0 & 0\\
        0 & 1 \\
    \end{pmatrix}
    ,
\end{align}
the $U(2)$ gauge symmetry is broken down to $U(1)_a\times U(1)_b$. 
In addition, there exists $M=m_a - m_b$ number of degenerate zero modes $\psi^{j,M}$ with $j=0,1,...,M-1$ for the bifundamental representation of $U(1)_a \times U(1)_b$ with the charge $(1,-1)$. Here and in what follows, we consider the positive magnetic flux $M>0$.

The gauge-invariant discrete translations on the degenerate chiral zero modes are generated by $e^{\frac{n_{y_1}}{M}D_{y_1}}$ and $e^{\frac{n_{y_2}}{M}D_{y_2}}$, where $n_{y_1}, n_{y_2} = 0,1,...,M-1$ and $D_{y_1,y_2}$ denote the covariant derivative. 
These discrete translations induce the following transformations of degenerate chiral zero modes:
\begin{align}
    Z^{n_{y_1}}\,&:\,e^{\frac{n_{y_1}}{M}D_{y_1}} \psi^{j,M}= e^{2\pi i\frac{j n_{y_1}}{M}} \psi^{j,M},
    \nonumber\\
    C^{n_{y_2}}\,&:\,e^{\frac{n_{y_2}}{M}D_{y_2}} \psi^{j,M}= \psi^{j+n_{y_2},M},
    \label{eq:ZnxCny}
\end{align}
i.e., a flavor symmetry in 4D low-energy effective field theory. 
Specifically, it results in the non-Abelian flavor symmetry $H_M\simeq (\mathbb{Z}_M \times \mathbb{Z}_M^{(Z)}) \rtimes \mathbb{Z}_M^{(C)}$, where 
$\mathbb{Z}_M^{(Z)}$ and $\mathbb{Z}_M^{(C)}$ are respectively described by $Z$ and $C$ in Eq.~(\ref{eq:ZnxCny}), and the other $\mathbb{Z}_M$ is given by 
$e^{\frac{2\pi i}{M}}{\rm diag}(1,1,\cdots,1)$ \cite{Abe:2009vi,Berasaluce-Gonzalez:2012abm,Marchesano:2013ega}.

It was known that the magnetized D-brane system is written in the operator formalism \cite{Cremades:2004wa,Abe:2014noa} which is useful to discuss the non-invertible symmetry. 
Indeed, the degenerate chiral zero modes are given by the ground state of the following Hamiltonian:
\begin{align}
    \hat{H} = \frac{1}{2} \hat{P}_1^2 + \frac{\omega^2}{2} \hat{Y}_1^2\,,
\label{eq:Harmonic}
\end{align}
where $\hat{Y}_1$ and $\hat{P}_1$ are given by the (real) coordinates of the torus $y_i$ and its momentum conjugate $p_i$:
\begin{align}
    \hat{Y}_1 &= \frac{\sqrt{2}}{\omega} \left( \hat{p}_2 - \frac{1}{2}F_{21}\hat{y}_1\right),
    \quad
    \hat{P}_1 = \sqrt{2}  \left( \hat{p}_1 - \frac{1}{2}F_{12}\hat{y}_2\right),
    \quad
    \hat{Y}_2 = -\frac{1}{2\pi M} \hat{T}_2,
    \quad
    \hat{P}_2 = \hat{T}_1,
\label{eq:YPdef}
\end{align}
with 
\begin{align}
\omega \equiv 2F_{12}
,\qquad
    \hat{T}_a \equiv \bm{e}_a^{\rm T}\left( \hat{\bm{p}} - \frac{1}{2}F^{\rm T}\hat{\bm{y}}\right)\,,
\end{align}
where we introduce the basis vector of torus $\bm{e}_a$ $(a=1,2)$. 
As in the quantum mechanics, we impose the canonical commutation relations:
\begin{align}
    [\hat{Y}_i, \hat{P}_j] = i \delta_{ij} \quad (i,j=1,2),
\label{eq:comrelation}
\end{align}
and otherwise 0. 
Since $\hat{Y}_2$ and $\hat{P}_2$ commute with the Hamiltonian, they are regarded as conserved quantities. 
It seems that $\hat{Y}_2$ and $\hat{P}_2$ are irrelevant to the dynamics of massless modes, but they give the following constraints:
\begin{align}
    e^{i\hat{P}_2}|\psi \rangle = |\psi \rangle,
    \qquad
    e^{2\pi i M \hat{Y}_2}|\psi \rangle = |\psi \rangle.
    \label{eq:constraints}
\end{align}
When we analyze these constraints in an eigenstate of $\hat{Y}_2$, i.e., $\hat{Y}_2|Y_2\rangle= Y_2|Y_2\rangle$, it turns out that the coordinate $Y_2$ is quantized as\footnote{For more details, see Ref.~\cite{Abe:2014noa}.}
\begin{align}
    Y_2 = \frac{j}{M},\quad j=0,1,...,M-1\,.
\end{align}
Then, the ground state of the Hamiltonian is spanned by the following orthonormal basis:
\begin{align}
    |\psi\rangle_{T^2}^{j,M} \equiv \left|\frac{j}{M}\right\rangle
    = e^{-i \frac{j}{M}\hat{P}_2}\left|0\right\rangle\,,
\end{align}
which parametrizes $M$ number of degenerate chiral zero modes. 
From
\begin{align}
    e^{-i\frac{1}{M}\hat{T}_2} |\psi\rangle_{T^2}^{j,M}
    &= e^{2\pi i \frac{j}{M}}\left|0\right\rangle\,,
    \nonumber\\
    e^{-i\frac{1}{M}\hat{T}_1} |\psi\rangle_{T^2}^{j,M}
    &= |\psi\rangle_{T^2}^{j+1,M}\,,
\end{align}
$\hat{T}_2$ and $\hat{T}_1$ respectively correspond to $Z$ and $C$ in Eq.~\eqref{eq:ZnxCny}, indicating that $\hat{T}_2$ and $\hat{T}_1$ are identified with the generators of flavor symmetry in the operator formalism.

We promote the above system to the $T^2/\mathbb{Z}_2$ orbifold. 
Since the $\mathbb{Z}_2$-twist flips the sign of basis vector of the torus $\bm{e}_a$, 
$\bm{e}_a \rightarrow -\bm{e}_a$ with $a=1,2$, the operator $\hat{T}_a$ also changes the sign under the $\mathbb{Z}_2$ twist: 
\begin{align}
\hat{T}_a \rightarrow -\hat{T}_a\,.
\end{align}
When we gauge the $\mathbb{Z}_2$ twist, the operators are restricted in the $\mathbb{Z}_2$-invariant form. 
Following Ref. \cite{Kobayashi:2024yqq}, the $\mathbb{Z}_2$-invariant operators are chosen as
\begin{align}
\hat{U}_{Y}^{(\lambda_Y)} \equiv e^{i \lambda_Y \hat{T}_2} + e^{- i \lambda_Y \hat{T}_2}\,,
\qquad
\hat{U}_{P}^{(\lambda_P)} \equiv e^{i\lambda_P \hat{T}_1} + e^{- i\lambda_P \hat{T}_1}\,,
\label{eq:UYUP}
\end{align}
with $\lambda_Y = n_Y/M$ and $\lambda_P=n_P/M$. 
Their fusion rules are given by
\begin{align}\label{eq:fusion-alg}
    \hat{U}_i^{(\lambda_i)}\hat{U}_i^{(\lambda_i^\prime)} &=\hat{U}_i^{(\lambda_i + \lambda_i^\prime)} + \hat{U}_i^{(\lambda_i - \lambda_i^\prime)}\,,
\end{align}
with $i=Y, P$. 
Note that the gauging \eqref{eq:UYUP} corresponds to the gauging the outer automorphism of each $\mathbb{Z}_M^{(Z)}$ and $\mathbb{Z}_M^{(C)}$.\footnote{Such a construction is useful to constrain the Yukawa matrices of quarks and leptons as pointed out in Ref. \cite{Kobayashi:2024cvp}.} Under this non-invertible symmetry, the $\mathbb{Z}_2$-even modes transform as
\begin{align}
   & \hat{U}^{(n_Y/M)}_Y\; :\; \left(e^{i\frac{n_Y}{M}\hat{T}^2}+e^{-i\frac{n_Y}{M}\hat{T}^2}\right)|\psi\rangle_{+}^{j,M}=2\mathrm{cos}\left(2\pi\frac{jn_Y}{M}\right) |\psi\rangle_{+}^{j,M},\nonumber\\
    &\hat{U}^{(n_P/M)}_P\; : \; \left(e^{i\frac{n_P}{M}\hat{T}_1}+e^{-i\frac{n_P}{M}\hat{T}_1}\right) |\psi\rangle_{+}^{j,M}= |\psi\rangle_{+}^{j+n_P,M}+ |\psi\rangle_{+}^{j-n_P,M},
\label{eq:Trf_generic}
\end{align}
with
\begin{align}\label{eq:wf-Z2}
    |\psi\rangle_{+}^{j,M} &= |\psi\rangle_{T^2}^{j,M} +  |\psi\rangle_{T^2}^{M-j,M} \,,\qquad
    j=1,2,...,\left\lfloor \frac{M}{2}\right\rfloor,
\end{align} 
up to the normalization factor. 
Here, $\lfloor \ast\rfloor$ denotes the floor function. The representation of matter zero modes under the above non-invertible symmetry was discussed in Ref. \cite{Kobayashi:2024yqq} which revealed that the $D_4$ symmetry was realized in the case of $M=$ even. 
Specifically, let us take $M=2p$ with $p\in \mathbb{Z}$. 
Then, the $\mathbb{Z}_2$-even modes are given by~\cite{Abe:2008fi,Abe:2013bca,Abe:2014noa}
\begin{align}
|\Psi\rangle_{\mathbb{Z}_2}^{M=2p}\equiv
    \begin{pmatrix}
        |\psi\rangle_{+}^{0,2p}\\
        |\psi\rangle_{+}^{j,2p}\\
        |\psi\rangle_{+}^{p,2p}\\
    \end{pmatrix}
    \,,
\end{align}
with $j=1,2,...,p-1$. 
The transformations of $\mathbb{Z}_2$-even modes under the non-invertible symmetry with $n_Y=n_P=p$ are given by\footnote{Here and in what follows, we rescale $\hat{U}_Y \rightarrow \hat{U}_Y/2$ and $\hat{U}_P \rightarrow \hat{U}_P/2$.}
\begin{align}
   \hat{U}_Y^{(1/2)}\,&:\,        
    |\Psi\rangle_{\mathbb{Z}_2}^{M=2p}
    \rightarrow 
        \begin{pmatrix}
        1 & 0 & \cdots & \cdots & \cdots &  0 & 0\\
        0 & -1 & \cdots & \cdots & \cdots & 0 & 0\\
        \vdots & \vdots &  &  &  & \vdots & \vdots \\
        0 & 0 & \cdots & \cos (\pi j) & \cdots & 0 & 0\\
        \vdots & \vdots &  &  &  & \vdots & \vdots \\
        0 & 0 & \cdots & \cdots & \cdots & -\cos(\pi p) & 0 \\
        0 & 0 & \cdots & \cdots & \cdots & 0 & \cos(\pi p) \\
    \end{pmatrix}
|\Psi\rangle_{\mathbb{Z}_2}^{M=2p}
    \,,
    \nonumber\\
    \hat{U}_P^{(1/2)}\,&:\,       
    |\Psi\rangle_{\mathbb{Z}_2}^{M=2p}
    \rightarrow 
        \begin{pmatrix}
        0  & 0 & \cdots &  0 & 1\\
        0  & 0 & \cdots &  1 & 0\\
        \vdots & \vdots & & \vdots & \vdots \\
        0 & 1 & \cdots  & 0 & 0 \\
        1 & 0 & \cdots & 0 & 0\\
    \end{pmatrix}
|\Psi\rangle_{\mathbb{Z}_2}^{M=2p}
    \,.
\label{eq:UYUP_trf}
\end{align}
It indicates that for $p=$ even, the pairs with the same $\mathbb{Z}_2^{(\hat{U}_Y)}$ charges are described by singlets, ${\bf 1}_{+\pm}$ and ${\bf 1}_{-\pm}$ of $D_4$. 
Here, two subscripts represent $\mathbb{Z}_2$ charges, i.e., 
even for $+$ and odd for $-$ under $\mathbb{Z}_2^{(\hat{U}_Y)}$ and $\mathbb{Z}_2^{(\hat{U}_P)}$. 
On the other hand, for $p=$ odd, the pairs with different $\mathbb{Z}_2^{(\hat{U}_Y)}$ charges are described by $D_4$ doublets. 
Table \ref{tab:D4-rep} shows $D_4$ representations of zero modes for several values of $M$, where $({\bf r})_{\rm odd}$ denotes the odd mode under the $\mathbb{Z}_2$ orbifold twist.

\begin{table}[ht]
\centering
\begin{tabular}{|c|c|} \hline
$M$ & Representations of $D_4$  \\ \hline \hline
2 & {\bf 2} \\
4 & ${\bf 1}_{++}$,  ${\bf 1}_{+-}$, ${\bf 1}_{-+}$,  $({\bf 1}_{--})_{\rm odd}$  \\
6 & $2\times {\bf 2}$, $({\bf 2})_{\rm odd}$ \\
8 & ${\bf 1}_{++}$,  ${\bf 1}_{+-}$, $3\times{\bf 1}_{-+}$, $3\times{\bf 1}_{--}$,  $(3\times {\bf 1}_{-+})_{\rm odd}$, $(3\times {\bf 1}_{--})_{\rm odd}$  \\
\hline
\end{tabular}
\caption{$D_4$ representations of zero modes.}
\label{tab:D4-rep}
\end{table}

So far, we have focused on $M$= even, but in the case of $M=$ odd, there is no "invertible" symmetry originating from $\mathbb{Z}_M^{(Z)}$ and $\mathbb{Z}_M^{(C)}$. 
However, even when $M=$ odd, the above non-invertible symmetry 
plays a role in determining the selection rules of allowed couplings \cite{Kobayashi:2024yqq}.

\subsection{Flavor symmetries in intersecting D-brane models}
\label{sec:flavor_IIA}

Flavor symmetries in magnetized D-brane models have 
been studied in Refs.~\cite{Abe:2009vi,Berasaluce-Gonzalez:2012abm,Marchesano:2013ega}.\footnote{See also for the coupling selection rules Ref.~\cite{Higaki:2005ie}.} 
Intersecting D-brane models and magnetized D-brane models are T-dual to each other \cite{Ibanez:2012zz}. 
Thus, in Ref.~\cite{Abe:2009vi}, the flavor symmetries were studied concretely in 
the magnetized D-brane side, but the flavor symmetries in the intersecting D-brane side are expected through the T-duality. 
Note that the 2D CFT picture is clear in the intersecting D-brane side, while it is not clear in the magnetized D-brane side. 
(See for 2D CFT in intersecting D-brane models Refs.~\cite{Cvetic:2003ch,Abel:2003vv}.) 
However, the effective field theoretic picture is clear in the magnetized D-brane side.

In this section, we first convert the discussions on the flavor symmetries in magnetized D-brane models \cite{Abe:2009vi} to ones in intersecting D-brane models. Then, we will study its non-invertible flavor symmetries following Ref.~\cite{Kobayashi:2024yqq}.

The open string between intersecting D-branes at angle $\theta$, has the following boundary condition:
\begin{align}
X(\sigma= \pi) -v=\theta (X(\sigma =0) -v),
\end{align}
where $v$ denotes the intersecting point of D-branes.
Note that we may have many intersecting points on $T^2$.
This boundary condition can be written as 
\begin{align}
X(\sigma= \pi)=\theta X(\sigma =0) +(1-\theta)v.
\end{align}

We choose the origin of our coordinates such that $v=0$ which is one of the intersecting points. 
Then, another intersecting point must correspond to the torus lattice point, $e$. 
The open strings with $v=0$ and $v=e$ must be identified.
Thus, open strings at intersecting points are classified by their boundary conditions $(\theta, (1-\theta)v)$, where the winding $ (1-\theta)v$ is classified mod $ (1-\theta)e$. Suppose that there are $M$ intersecting points on $T^2$. Then, we can write $v=ek/M$ ($k=0,\cdots, M-1$), 
indicating that there are $M$ kind of open strings with the winding number 
$(1-\theta)ke/M$ ($k=0,\cdots, M-1$).
Note that the winding number of intersecting D-brane corresponds to discrete momenta in the T-dual side, i.e. the magnetized D-brane models.
Because of the modulo structure of winding numbers, 
there appears the $\mathbb{Z}_M$ symmetry and each open string has 
different charge. That corresponds to $\mathbb{Z}_M^{(Z)}$ in Ref.~\cite{Abe:2009vi}.

We have chosen one of intersecting points as the origin, i.e. $v=0$.
However, we may choose another intersecting point as the origin.
That corresponds to the $\mathbb{Z}_M$ permutation, i.e. 
$\mathbb{Z}_M^{(C)}$ in Ref.~\cite{Abe:2009vi}.
They make the total flavor symmetry, $(\mathbb{Z}^{(Z)}_M\times \mathbb{Z}_M) \rtimes \mathbb{Z}^{(C)}_M$.
 
Let us introduce the twisted fields $\sigma_{\theta,v}$ to write 2D CFT.
The twisted field $\sigma_{\theta,v}$ creates the ground state of open string with the boundary condition 
$(\theta, (1-\theta)v)$ from the untwisted ground state.
The twisted fields have representations of the above flavor symmetry, $(\mathbb{Z}^{(Z)}_M\times \mathbb{Z}_M) \rtimes \mathbb{Z}^{(C)}_M$.
That is, $\mathbb{Z}^{(Z)}_M$ is represented by
\begin{align}
U_X=\begin{pmatrix}
1 & & & \\
& e^{2\pi i/M} & & \\
& & \ddots & \\
& & & e^{2\pi i(M-1)/M}
\end{pmatrix},
\end{align}
on the twisted fields $(\sigma_{\theta,0}, \sigma_{\theta, e/M},\cdots, \sigma_{\theta, (M-1)e/M})^T$, and 
$\mathbb{Z}^{(C)}_M$ is represented by
\begin{align}
U_P=\begin{pmatrix}
0 &1 &\cdots & \\
0& 0 &1 &\cdots \\
0&\cdots & \cdots & \cdots\\
1 & \cdots & \cdots  & \cdots
\end{pmatrix}.
\end{align}
They respectively correspond to $Z$ and $C$ in Eq.~\eqref{eq:ZnxCny} with $n_{y_1}=n_{y_2}=1$.

The transformation of $U_X$ is generated by the following operator:
\begin{align}
    \hat U_X=e^{2\pi i /(eM)\int d\sigma \partial X},
\end{align}
because the boundary condition of string $X(\sigma = \pi)$ is shifted by the insertion of the twisted fields like the monodromy structure due to twisted fields \cite{Hamidi:1986vh,Dixon:1986qv} and topological defect lines \cite{Chang:2018iay}.
A similar operator is also introduced to represent the non-invertible symmetry  on $S^1/\mathbb{Z}_2$ in Ref.~\cite{Heckman:2024obe}.
The above operator corresponds to $e^{i\hat T_2}$ in the magnetized one.
After gauging $\mathbb{Z}_2$, the symmetry operator is constrained in the $\mathbb{Z}_2$-invariant way and written as 
\begin{align}
\hat {U}_X^{'} \equiv \hat U_X + \hat U_{-X},
\end{align}
similar to the magnetized one. 
We can introduce the operator $\hat U_P$ for $U_P$ by using the momentum $P$ conjugate to $X$ as $\hat U_P=e^{i P}$.
Since $X$ and $P$ satisfy the same commutation relation as $\hat Y_2$ and $\hat P_2$, $\hat U_P$ shifts the boundary condition of $X$.
Similar to $\hat{U}_X$, we can construct the $\mathbb{Z}_2$-invariant operator 
$\hat {U}_P^{'}\equiv \hat U_P + \hat U_{-P}$.
It results in the same discussion in the previous section, i.e., $D_4$ symmetry for the twisted fields in the case of $M=$ even. 
We conclude that non-invertible symmetries between magnetized D-brane models and intersecting D-brane models 
are equivalent because of T-duality.
We can use one of the magnetized and intersecting D-brane pictures, which is convenient to study.

We would study more details of non-invertible symmetries in 2D CFT for intersecting D-brane models elsewhere.
At any rate, intersecting D-brane models are the T-dual to magnetized D-brane models.
In what follows, we study quantum aspects of non-invertible symmetries within the framework of magnetized compactifications.

\subsection{Loop effects}
\label{sec:IIB_loop}

Intersecting/magnetized D-brane models 
lead to the same symmetry as shown in the previous section.
We study perturbative effects on these non-invertible symmetry 
in the terminology of magnetized compactifications. 
Before $\mathbb{Z}_2$ orbifolding, 
fields affected by the magnetic flux $M$ have the 
flavor symmetry $H_M\simeq (\mathbb{Z}_M \times \mathbb{Z}_M^{(Z)}) \rtimes \mathbb{Z}_M^{(C)}$.
A model includes various fields with different fluxes $M_i$.
The common symmetry $H_m$ remains, where $m$ is the greatest common divisor of $M_i$, although 
no symmetry remains for $m=1$.
The bosonic and fermionic modes have the same wave functions and the same quantum numbers under $H_m$.
We denote them by the same letter $\psi^{i,M}$.
The coupling terms  $\psi^{k_1,M_1} \cdots \psi^{k_N,M_N}$ in the bare Lagrangian are allowed if the product of their $\mathbb{Z}_m^{(Z)}$ satisfies 
\begin{align}
\label{eq:selec-rule-ZM}
    a_{k_1}\cdots a_{k_N} = e,
\end{align}
where $a_{k_i}$ denotes the $\mathbb{Z}_m^{(Z)}$ element of each mode $\psi^{k_iM}$, i.e., 
$a_{k_i}=e^{2\pi i k_i/m}$.

Now, we study the coupling selection rules after $\mathbb{Z}_2$ orbifolding.
When $m=2$, $D_4$ symmetry remains after orbifolding.
The discussion in Sec.~\ref{sec:generic-loop} implies that this $D_4$ symmetry is still valid when we take into account of loop effects.
When $m>2$, the selection rules due to other non-invertible symmetries 
remain in the bare Lagrangian
after orbifolding.
However, they are violated by loop effects.
We examine these aspects explicitly.

On the $T^2/\mathbb{Z}_2$ orbifold, the modes $\psi^{i,M}$ are decomposed to the $\mathbb{Z}_2$ even modes,
\begin{align}
    \psi_+^{i,M}=\psi^{i,M} + \psi^{M-i,M},
\end{align}
and odd modes,
\begin{align}
    \psi_-^{i,M}=\psi^{i,M} - \psi^{M-i,M},
\end{align}
up to normalization.
Normalization factors are irrelevant to 
our discussions and we omit them.
Let us focus on models including only $\mathbb{Z}_2$-even modes.
The following discussions are applicable 
even when the model includes $\mathbb{Z}_2$-odd modes.
Note that they have different parities under the unbroken $\mathbb{Z}_2$ (invertible) symmetry.
Because of the above liner combinations, the even modes as well as the odd modes correspond to different 
$\mathbb{Z}_M^{(Z)}$ and $\mathbb{Z}_m^{(Z)}$ charges,
i.e., $a_k=e^{2\pi ik}$ and $a_{m-k}=e^{2\pi i(m-k)}$, 
where $M=mn$, although $k=m-k$ (mod $m$) in certain modes, e.g. $k=0$ and $k=m/2$ (mod $m$).
These possible charges to a single mode $\psi_+^{i,M}$ as well as 
$\psi_-^{i,M}$ are the key points to violate the selection rules due to loop effects in Sec.~\ref{sec:generic-loop}.
We denote the set including $a_k=e^{2\pi ik}$ and $a_{m-k}=e^{2\pi i(m-k)}$ 
by $[a_k]$, which corresponds to a single mode $\psi_+^{k,M}$.
Then, the selection rules on $T^2/\mathbb{Z}_2$ is explained by terminology of $[a_k]$ as follows.

\paragraph{Selection rule:}
The coupling term $\psi^{k_1,M_1}\cdots \psi^{k_N,M_N}$ is allowed in the 
bare Lagrangian only if one can choose either $a_{k_i}$ or $a_{m-k_i}$ from  $[a_{k_i}]$, which satisfy the selection rule (\ref{eq:selec-rule-ZM}).

\paragraph{}
This selection rule is still valid at tree-level diagrams, but it can be violated 
by loop effects as generic discussions in Sec.~\ref{sec:generic-loop}.
The $D_4$ symmetry still remains even when we take account of loop effects.
The key point is that one mode $\psi^{k,M}_+$ with $[a_k]$ corresponds to two possible representations $a_k=e^{2\pi ik}$ and $a_{m-k}=e^{2\pi i(m-k)}$ of $\mathbb{Z}_m^{(Z)}$. 
In the internal lines, we insert $\psi^{k,M}(\psi^{k,M})^*$, which corresponds to $[a_k][a_{-k}]$.
Two possibilities in each of $[a_k]$ and $[a_{-k}]$ can violate the above selection rule in loop diagrams.
We examine these aspects by using concrete 
examples in what follows.

\subsubsection{Model with $m=2$}

We start with the model with $m=2$, which has the $D_4$ symmetry.
When $M=2$, we have two modes, $\psi^{0,M}$ and $\psi^{1,M}$.
Each of them corresponds to a single representation, $a_0=1$ or $a_1=-1$.
In the internal line corresponding to $[a_k][a_{-k}]$, there is no ambiguity.
That is, $[a_0][a_{-0}]$ and $[a_1][a_{-1}]$ correspond to uniquely  
$a_0a_0$ and $a_1a_1$, respectively.
These modes correspond to the $D_4$ doublet \cite{Abe:2009vi}.

Next, we study the modes with $M=4$.
We have three $\mathbb{Z}_2$ even modes,
\begin{align}
\label{eq:Z2-even-M4}
    \psi_+^{0,4}=\psi^{0,4}, \qquad \psi_+^{2,4}=\psi^{2,4}, \qquad 
    \psi_+^{1,4}=\psi^{1,4} + \psi^{3,4}.
\end{align}
Two $\mathbb{Z}_4^{(Z)}$ charges are mixed in $\psi_+^{1,4}$.
The insertion of $\psi_+^{1,4}(\psi_+^{1,4})^*$ in internal lines 
can change the $\mathbb{Z}_4^{(Z)}$ by 1 and 3 (mod 4).
Thus, the selection ruled due to $\mathbb{Z}_4^{(Z)}$ is violated by loop effects.
However, the common symmetry is $\mathbb{Z}_{m=2}^{(Z)}$.
Each of the above three modes definitely has a single representation of 
$\mathbb{Z}_{m=2}^{(Z)}$.
Indeed, $\psi^{1,4}$ correspond to 
${\bf 1}_{-+}$ of $D_4$ in the notation of Ref.~\cite{Abe:2009vi}, 
while $\psi^{0,4} + \psi^{2,4}$ and $\psi^{0,4} - \psi^{2,4}$ correspond to 
${\bf 1}_{++}$ and ${\bf 1}_{+-}$ of $D_4$, respectively.
We find a one-to-one correspondence between $D_4$ representations and 
modes including liner combinations.
Therefore, the selection rule due to $D_4$ remains valid 
through loop effects.
We can discuss the modes with larger $M=2n$ ($n>2$).
They also correspond to $D_4$ representations \cite{Abe:2009vi}.
Thus, the $D_4$ symmetry remain when $m=2$.

\subsubsection{Model with $m=3$}

Here we study the model with $m=3$.
For $M=3$, we have two $\mathbb{Z}_2$ even modes:
\begin{align}
    \psi^{0,3}_+=\psi^{0,3},\qquad \psi^{1,3}_+=\psi^{1,3}+\psi^{2,3}.
\end{align}
In the bare Lagrangian, the above selection rule due to $\mathbb{Z}_3^{(Z)}$ is valid.
However, the insertion of $\psi^{1,3}_+(\psi^{1,3}_+)^*$ in the internal line change the $\mathbb{Z}_3^{Z}$ charge by 0, 1, 2 (mod 3).
Then, the selection rule due to $\mathbb{Z}_3^{Z}$ has no meaning 
in loop diagram, while it is still valid in the tree-level diagram.

\subsubsection{Model with $m>3$}

Similarly, we can discuss the models with $m>3$.
For example, in the model with $m=4$, the $\mathbb{Z}_2$-even modes with $M=4$ are shown in Eq.~(\ref{eq:Z2-even-M4}).
We have already shown that the insertion of $\psi^{1,4}_+(\psi^{1,4}_+)^*$
in the internal line changes the $\mathbb{Z}_4^{(Z)}$ charge by 
0, 1, 3.
The selection rule due to $\mathbb{Z}_4^{(C)}$ is violated by loop effects.
However, the $D_4$ symmetry remains.
Similarly, when $m=$ even, the $D_4$ symmetry remains, but other symmetries are violated by loop effects.
When $m=$ odd, all the selection rules are violated by loop effects.

So far, we have studied the $\mathbb{Z}_2$-even modes.
We can also discuss the models including $\mathbb{Z}_2$-odd modes.
For example, when $M=4$, the $\mathbb{Z}_2$ odd mode $\psi_-^{1,4}=\psi^{1,4} - \psi^{3,4}$
corresponds to ${\bf 1}_{--}$ of $D_4$.
Since it corresponds to a single representation of $D_4$, 
the $D_4$ symmetry is not violated by loop effects.
Similarly, when $M=$ even, the $\mathbb{Z}_2$ odd modes correspond to 
unique representations of $D_4$.
Thus, the $D_4$ symmetry remains even when we include the $\mathbb{Z}_2^{(Z)}$-odd modes for $M=$ even.

We have found that the $D_4$ symmetry remains for $m=2n$, even when we take into account loop effects.
In the next section, non-perturbative effects on the $D_4$ symmetry will be studied.

\section{Non-invertible symmetries in D-brane models at non-perturbative level}
\label{sec:np}

In this section, we discuss the fate of non-invertible symmetry at the non-perturbative level in the context of type IIB magnetized D-brane models. 
D-brane instanton effects are calculable non-perturbative effects, although there may be other non-perturbative effects.
In this section, we examine D-brane instanton effects 
on the non-invertible symmetry by studying possible D-brane instanton configurations.
In Sec. \ref{sec:2-point}, we examine the symmetry of the Majorana masses of right-handed neutrinos by use of an illustrating example. 
Similarly, we study an example of the 3-point couplings 
of singlet fields in Sec. \ref{sec:3-point}. 
A generic $n$-point coupling of singlet fields under the Standard Model (SM) gauge group is studied in Sec. \ref{sec:n-point}. 
Finally, we discuss the symmetry of $\mu$-term as a concrete example in Sec. \ref{sec:mu-term} in which the fields have representations under the SM gauge group.

\subsection{2-point couplings}
\label{sec:2-point}

In this section, we study whether the $D_4$ symmetry remaining at perturbative level is broken by D-brane instanton effects. 
When D-brane instantons appear, new charged zero modes $\beta_i$ and $\gamma_i$ are induced.
They may have non-vanishing couplings with matter fields.
Such couplings induce correction terms in Lagrangian of low-energy effective field theory after we integrate charged zero modes, $\beta_i$ and $\gamma_i$.

It would be better to start with concrete examples.
First, we study right-handed neutrino mass terms 
which are induced possibly from~\cite{Blumenhagen:2006xt,Ibanez:2006da}:
\begin{align}
e^{-S_{\rm cl}}\int d^2 \beta d^2\gamma e^{-\sum_{ija} d^{i,j,a}_{T^2/\mathbb{Z}_2}\beta_i\gamma_j N_a},
\end{align}
where $S_{\rm cl}$ denotes the instanton action, which depends on moduli, 
$\beta_i, \gamma_j$ are charged zero modes, which are induced by D-brane instanton, 
$N_a$ are right-handed neutrinos, and $d^{i,j,a}_{T^2/\mathbb{Z}_2
}$ are their 3-point couplings on $T^2/\mathbb{Z}_2$. 
(See also Refs.~\cite{Ibanez:2007rs,Antusch:2007jd,Kobayashi:2015siy,Hoshiya:2021nux} for D-brane instanton calculations in explicit models.) 
Since we are interested in the invariance of $D_4$ symmetry at the non-perturbative level, we focus on the case where the magnetic flux of $N_a$ is even which we label $M_N=2N$ in the following analysis. 
To induce $n$-point couplings of $N_a$, we require the same number of charged zero modes for $\beta_i$ and $\gamma_j$. 
Taking into account $M_{\beta_i} + M_{\gamma_j}= M_N$, the value of magnetic fluxes are uniquely chosen as $M_\beta=M_\gamma=N$. 
This is because the number of zero modes has the mod 2 structure as in Table~\ref{tab:zeromodes} where we present the case including Scherk-Schwarz (SS) phases $(\alpha_1, \alpha_\tau)$ \cite{Abe:2013bca}. 
In particular, we focus on $\mathbb{Z}_2$-even modes without SS phases for $N_a$. We refer to the model with $(N_\beta,N_\gamma, N_a)$ as 
the $N_\beta-N_\gamma-N_a$ model.

\begin{table}[ht]
\centering
\begin{tabular}{|c||c|c|c|c|c|c|c|c|c|c|c|c|c||c|} \hline
& $M$ & 1 & 2 & 3 & 4 & 5 & 6 & 7 & 8 & 9 & 10 & 11 & 12 & $N_{\rm gen}$ \\ \cline{2-15}
$(\alpha_1, \alpha_{\tau}) = (0,0)$ &  $\mathbb{Z}_2$-even & 1 & 2 & 2 & 3 & 3 & 4 & 4 & 5 & 5 & 6 & 6 & 7 & $\lfloor \frac{M}{2}\rfloor$ +1\\ \cline{2-15}
& $\mathbb{Z}_2$-odd & 0 & 0 & 1 & 1 & 2 & 2 & 3 & 3 & 4 & 4 & 5 & 5 & $\lfloor \frac{M+1}{2}\rfloor$ -1 \\ 
\hline \hline
& $M$ & 1 & 2 & 3 & 4 & 5 & 6 & 7 & 8 & 9 & 10 & 11 & 12 & $N_{\rm gen}$\\ \cline{2-15}
$(\alpha_1, \alpha_{\tau}) = (1/2,0)$  & $\mathbb{Z}_2$-even & 1 & 1 & 2 & 2 & 3 & 3 & 4 & 4 & 5 & 5 & 6 & 6 & $\lfloor \frac{M+1}{2}\rfloor$\\ \cline{2-15}
& $\mathbb{Z}_2$-odd & 0 & 1 & 1 & 2 & 2 & 3 & 3 & 4 & 4 & 5 & 5 & 6 & $\lfloor \frac{M}{2}\rfloor$ \\ 
\hline \hline
& $M$ & 1 & 2 & 3 & 4 & 5 & 6 & 7 & 8 & 9 & 10 & 11 & 12 & $N_{\rm gen}$\\ \cline{2-15}
 $(\alpha_1, \alpha_{\tau}) = (0,1/2)$ & $\mathbb{Z}_2$-even & 1 & 1 & 2 & 2 & 3 & 3 & 4 & 4 & 5 & 5 & 6 & 6 & $\lfloor \frac{M+1}{2}\rfloor$ \\ \cline{2-15}
& $\mathbb{Z}_2$-odd & 0 & 1 & 1 & 2 & 2 & 3 & 3 & 4 & 4 & 5 & 5 & 6 & $\lfloor \frac{M}{2}\rfloor$ \\ \hline \hline
& $M$ & 1 & 2 & 3 & 4 & 5 & 6 & 7 & 8 & 9 & 10 & 11 & 12 & $N_{\rm gen}$\\ \cline{2-15}
$(\alpha_1, \alpha_{\tau}) = (1/2,1/2)$ & $\mathbb{Z}_2$-even & 0 & 1 & 1 & 2 & 2 & 3 & 3 & 4 & 4 & 5 & 5 & 6 & $\lfloor \frac{M}{2}\rfloor$ \\ \cline{2-15}
& $\mathbb{Z}_2$-odd & 1 & 1 & 2 & 2 & 3 & 3 & 4 & 4 & 5 & 5 & 6 & 6 & $\lfloor \frac{M+1}{2}\rfloor$ \\ \hline 
\end{tabular}
\caption{The number of zero-modes $N_{\rm gen}$ with the SS phases $(\alpha_1, \alpha_{\tau}) = (0,0), (1/2, 0), (0, 1/2), (1/2,1/2)$.}
\label{tab:zeromodes}
\end{table}

When we introduce the SS phases $(\alpha_1, \alpha_\tau)$, the boundary conditions of fields $\psi_{T^2}$ with $U(1)$ charge $1$ are given by
\begin{align}
    \psi_{T^2}(z+1) &= e^{2\pi i \alpha_1} e^{\pi i M \frac{{\rm Im}(z)}{{\rm Im}(\tau)}} \psi_{T^2}(z),
    \nonumber\\
    \psi_{T^2}(z+\tau) &= e^{2\pi i \alpha_\tau} e^{\pi i M \frac{{\rm Im}(\Bar{\tau}z)}{{\rm Im}(\tau)}} \psi_{T^2}(z),
\end{align}
where we consider the positive magnetic flux $M>0$. 
Then, the wavefunction of chiral zero modes on $T^2$ is written as
\begin{align}
    \psi_{T^2}^{(i+\alpha_1^{(i)}, \alpha_\tau^{(i)}),M}(z) = \left(\frac{M}{{\cal A}^2}\right)^{1/4}
    e^{2\pi i \frac{(i+\alpha_1^{(i)})\alpha^{(i)}_\tau}{M}} e^{\pi i Mz \frac{{\rm Im}(z)}{{\rm Im}(\tau)}}
\vartheta\left[
    \begin{array}{c}
        \frac{i+\alpha_1^{(i)}}{M}\\
        -\alpha_\tau^{(i)}        
    \end{array}
    \right](Mz,M \tau),
\end{align}
where $\vartheta$ denotes the Jacobi-Theta function:
\begin{align}
        \vartheta\left[\begin{array}{cc}
         a  \\
         b 
    \end{array}\right](\nu,\tau)
    =\sum_{\ell\in\mathbb{Z}}e^{\pi i(a+\ell)^2\tau}e^{2\pi i(a+\ell)(\nu+b)}.
\end{align}
A linear combination gives rise to the $\mathbb{Z}_2$-even and -odd modes on $T^2/\mathbb{Z}_2$:
\begin{align}
    \psi_+^{(j+\alpha_1^{(j)}, \alpha_\tau^{(j)}),M} (z)&=\psi_{T^2}^{(j+\alpha_1^{(j)}, \alpha_\tau^{(j)}),M}(z) + e^{-4\pi i (j+\alpha_1^{(j)})\alpha_\tau^{(j)}/M} \psi_{T^2}^{(M-(j+\alpha_1^{(j)}), \alpha_\tau^{(j)}),M}(z),
    \nonumber\\
    \psi_-^{(j+\alpha_1^{(j)}, \alpha_\tau^{(j)}),M}(z)&=\psi_{T^2}^{(j+\alpha_1^{(j)}, \alpha_\tau^{(j)}),M}(z) - e^{-4\pi i (j+\alpha_1^{(j)})\alpha_\tau^{(j)}/M} \psi_{T^2}^{(M-(j+\alpha_1^{(j)}), \alpha_\tau^{(j)}),M}(z),
\end{align}
up to the normalization factor. 
Note that the index $j$ runs over $j=0,1,\dots j_{\rm max}$ 
with $j_{\rm max} = N_{\rm gen}-1$ where $N_{\rm gen}$ is the number of 
linearly-independent zero modes as shown in Table \ref{tab:zeromodes}. 

When we introduce the SS phases, the operators $\{\hat{T}_1, \hat{T}_2\}$ are modified as 
$\{\hat{T}_1 - 2\alpha_1, \hat{T}_2 - 2\alpha_\tau\}$, according to which the transformations of matter fields under the non-invertible symmetry are given by
\begin{align}
    |\psi\rangle_{+}^{(i+\alpha_1^{(i)}, \alpha_\tau^{(i)}),M=2p}
    &\xrightarrow{\hat{U}_Y^{(1/2)}} 
    (-1)^{(i-2\alpha_1)}
    |\psi\rangle_{+}^{(i+\alpha_1^{(i)}, \alpha_\tau^{(i)}),M=2p}
    \,,
    \nonumber\\
    |\psi\rangle_{+}^{(i+\alpha_1^{(i)}, \alpha_\tau^{(i)}),M=2p}
    &\xrightarrow{\hat{U}_P^{(1/2)}} 
   \delta_{i+k, i_{\rm max}}
    |\psi\rangle_{+}^{(k+\alpha_1^{(k)}, \alpha_\tau^{(k)}),M=2p}
    \,.
\label{eq:UYUP_trf_SS}
\end{align}

The explicit form of 3-point coupling $d^{i,j,a}_{T^2/\mathbb{Z}_2}$ in the $N-N-2N$ model is defined by
\begin{align}
    \nonumber  d^{i,j,a}_{T^2/\mathbb{Z}_2} &=\int_{T^2/\mathbb{Z}_2}dzd\Bar{z}\,\psi_+^{(i+\alpha_1^{(i)}, \alpha_\tau^{(i)}),N}\cdot \psi_+^{(j+\alpha_1^{(j)}, \alpha_\tau^{(j)}),N}\cdot\left(\psi_+^{(a+\alpha_1^{(a)}, \alpha_\tau^{(a)}),2N}\right)^*\\
   \nonumber &=d^{i,j,a}_{T^2}+(u^{(a)})^{1/2}d^{i,j,2N-a}_{T^2}+u^{(j)}d^{i,N-j,a}_{T^2}+u^{(j)}(u^{(a)})^{1/2}d^{i,N-j,2N-a}_{T^2}+u^{(i)}d^{N-i,j,a}_{T^2}\\
    &\quad+u^{(i)}(u^{(a)})^{1/2}d^{N-i,j,2N-a}_{T^2}+u^{(i)}u^{(j)}d^{N-i,N-j,a}_{T^2}+u^{(i)}u^{(j)}(u^{(a)})^{1/2}d^{N-i,N-j,2N-a}_{T^2},
\label{eq:dija_Z2_SS}
\end{align}
with
\begin{align}
    u^{(i)}=e^{-4\pi i(i+\alpha^{(i)}_1)\alpha^{(i)}_\tau/N},
    \quad
    u^{(j)}=e^{-4\pi i(j+\alpha^{(j)}_1)\alpha^{(j)}_\tau/N},
    \quad
    u^{(a)}=e^{-4\pi i(a+\alpha^{(a)}_1)\alpha^{(a)}_\tau/(2N)},
\end{align}
where the range of $i,j,a$ is respectively given by $i=0,1,\dots i_{\rm max}$, $j=0,1\dots j_{\rm max}$, $a=0,1,\dots a_{\rm max}$.  
Then, we rewrite the 3-point couplings of $\mathbb{Z}_2$-even modes on those of wavefunctions on $T^2$:
\begin{align}
    d^{i,j,a}_{T^2} &=\int_{T^2}dz d\Bar{z}\psi_{T^2}^{(i+\alpha_1^{(i)}, \alpha_\tau^{(i)}),M_i}(z)\cdot\psi_{T^2}^{(j+\alpha_1^{(j)}, \alpha_\tau^{(j)}),M_j}(z)\cdot\left(\psi_{T^2}^{(a+\alpha_1^{(a)}, \alpha_\tau^{(a)}),M_a}(z)\right)^*
    \nonumber\\
    &=c_{(M_i-M_j-M_a)}
    \exp{2\pi i \left(\frac{(i+\alpha_1^{(i)})\alpha_\tau^{(i)}}{M_i} + \frac{(j+\alpha_1^{(j)})\alpha_\tau^{(j)}}{M_j} - \frac{(a+\alpha_1^{(a)})\alpha_\tau^{(a)}}{M_a} \right)}
    \nonumber\\
    &\times \sum_{m=0}^{M_a-1}\vartheta\left[
    \begin{array}{c}
        \frac{M_j(i+\alpha_1^{(i)})-M_i(j+\alpha_1^{(j)}) +M_iM_j m}{M_iM_jM_a}\\
        0        
    \end{array}
    \right](M_i\alpha_\tau^{(i)} - M_j\alpha_\tau^{(i)} ,M_iM_jM_a \tau)
    \nonumber\\
    &\times \delta_{(i+\alpha_1^{(i)}) +(j+\alpha_1^{(j)})-(a+\alpha_1^{(a)}),M_a \ell-M_im},
\label{eq:dija_SS}
\end{align}
with $\ell \in \mathbb{Z}$, $M_i + M_j = M_a$, $\alpha_1^{(i)} + \alpha_1^{(j)} = \alpha_1^{(a)}$, $\alpha_\tau^{(i)} + \alpha_\tau^{(j)} = \alpha_\tau^{(a)}$ and $c_{(M_i-M_j-M_a)}$ being a normalization factor:
\begin{align}
    c_{(M_i-M_j-M_a)}=(2\,{\rm Im}(\tau))^{-1/2} {\cal A}^{-1/2}\left|\frac{M_iM_j}{M_a}\right|^{1/4}.
\end{align}
Here, ${\cal A}$ stands for the area of $T^2$.

If the number of charged zero modes is equal to two for each of $\beta_i$ and $\gamma_i$, by integrating out the zero modes, one can obtain 2-point couplings: 
\begin{align}
e^{-S_{\mathrm{cl}}}\sum_{i,j,k,\ell,a,b}(\varepsilon_{ij}\varepsilon_{k\ell} d^{ik}_ad^{j\ell}_b)N_aN_b,
\end{align}
which is the right-handed neutrino mass terms, i.e.,
\begin{align}
    M_{ab} \equiv e^{-S_{\mathrm{cl}}}m_{ab},
\end{align}
with
\begin{align}
    m_{ab} \equiv \sum_{i,j,k,\ell}\varepsilon_{ij}\varepsilon_{k\ell} d^{ik}_ad^{j\ell}_b.
\end{align}
In the following sections, we focus on concrete magnetized D-brane models on $T^2/\mathbb{Z}_2$ and study the transformation of matter fields under the non-invertible transformations generated by Eq.~\eqref{eq:UYUP}. 
In particular, we show an example leading to the right-handed neutrino mass terms for $M_\beta=M_\gamma=2$ and $M_a=4$.

\paragraph{2-2-4 model}\,\\

Let us start with the case where the right-handed neutrinos have $M_N=4$ and vanishing SS phases, which leads to three $\mathbb{Z}_2$ even zero modes. Indeed, among four zero modes on $T^2$, the wavefunctions of $\mathbb{Z}_2$-even modes  are described by the following three modes:
\begin{align}
|\Psi\rangle_{\mathbb{Z}_2, +}^{M=4}\equiv
\left(
\begin{array}{c}
    |\psi\rangle^{0,4}_{+}
    \\
    |\psi\rangle^{1,4}_{+}
    \\    
    |\psi\rangle^{2,4}_{+} 
\end{array}
\right)
=
\left(
\begin{array}{c}
    |\psi\rangle^{0,4}_{T^2}
    \\
   \frac{1}{\sqrt{2}}\left(|\psi\rangle^{1,4}_{T^2} + |\psi\rangle^{3,4}_{T^2}\right)
    \\    
   |\psi\rangle^{2,4}_{T^2}
\end{array}
\right),
\end{align}
and their transformations under the non-invertible symmetry are given by
\begin{align}
    \hat{U}_Y^{(1/2)}\,&:\,        
    |\Psi\rangle_{\mathbb{Z}_2, +}^{M=4}
    \rightarrow 
        \begin{pmatrix}
        1 & 0 & 0\\
        0 & -1 & 0\\
        0 & 0 & 1 \\
    \end{pmatrix}
    |\Psi\rangle_{\mathbb{Z}_2, +}^{M=4}
    \,,
    \nonumber\\
   \hat{U}_P^{(1/2)}\,&:\,       
    |\Psi\rangle_{\mathbb{Z}_2, +}^{M=4}
    \rightarrow 
        \begin{pmatrix}
        0  & 0 & 1\\
        0  & 1 & 0\\
        1 & 0 & 0\\
    \end{pmatrix}
    |\Psi\rangle_{\mathbb{Z}_2, +}^{M=4}
    \,.
\label{eq:M=4_even_trf}
\end{align}
Note that both the representations $\hat{U}_Y^{(1/2)}$ and $\hat{U}_P^{(1/2)}$ lead to $\mathbb{Z}_2^{(\hat{U}_Y)}$ and $\mathbb{Z}^{(\hat{U}_P)}_2$ symmetries 
under which $|\psi\rangle_{T^2}^{0,4}$ and $|\psi\rangle_{T^2}^{2,4}$ are singlets ${\bf 1}_{++}$ and ${\bf 1}_{+-}$. 
The representation of the other mode $|\psi\rangle_{T^2}^{1,4}$ is ${\bf 1}_{-+}$. 
Here, $\mathbb{Z}_2^{(\hat{U}_Y)} \times \mathbb{Z}^{(\hat{U}_P)}_2$ is a subgroup of $D_4$, which contains four singlets: ${\bf 1}_{++}, {\bf 1}_{+-}, {\bf 1}_{-+}, {\bf 1}_{--}$. 
(For more details of the $D_4$ representations, see, e.g., Refs.~\cite{Ishimori:2010au,Kobayashi:2022moq}.)

When $M_N=4$, there is a unique D-brane configuration where the magnetic fluxes of the charged zero modes $\beta_i$ and $\gamma_j$ are $M_\beta = M_\gamma =2$. The number of the charged zero modes for each of $\beta_i$ and $\gamma_j$ must be two in order to induce the mass terms.
We find the unique possibility that $\beta_i$ and $\gamma_j$ correspond to the $\mathbb{Z}_2$-even modes with vanishing SS phase as shown in Table \ref{tab:zeromodes}.
Their explicit form of $\mathbb{Z}_2$-even wavefunctions is
\begin{align}
|\Psi\rangle_{\mathbb{Z}_2,+}^{M=2}\equiv
\left(
\begin{array}{c}
    |\psi\rangle^{0,2}_{+}\\
    |\psi\rangle^{1,2}_{+} 
\end{array}
\right)
=
\left(
\begin{array}{c}
    |\psi\rangle^{0,2}_{T^2}\\
    |\psi\rangle^{1,2}_{T^2}
\end{array}
\right),
\label{eq:M=2_even_wave}
\end{align}
whose transformations under the non-invertible symmetry are given by
\begin{align}
   \hat{U}_Y^{(1/2)}\,&:\,       
        |\Psi\rangle_{\mathbb{Z}_2,+}^{M=2}
    \rightarrow 
    \begin{pmatrix}
        1 & 0\\
        0 & -1
    \end{pmatrix}
        |\Psi\rangle_{\mathbb{Z}_2,+}^{M=2}
    ,
    \nonumber\\
   \hat{U}_P^{(1/2)}\,&:\,       
        |\Psi\rangle_{\mathbb{Z}_2,+}^{M=2}
    \rightarrow 
    \begin{pmatrix}
        0 & 1\\
        1 & 0
    \end{pmatrix}
       |\Psi\rangle_{\mathbb{Z}_2,+}^{M=2}
    .
\label{eq:M=2_even_trf}
\end{align}
The above representations obey
\begin{align}
    &\hat{U}_{Y}^{(1/2)}= \sigma_3\,, \quad 
    \hat{U}_{P}^{(1/2)}= \sigma_1\,,
    \quad
    \hat{U}_{Y}^{(1)} = \hat{U}_{P}^{(1)} = \mathbb{I}_2\,,
\end{align}
where we introduce the Pauli matrices $\sigma_{1,3}$. 
Since $\hat{U}_Y^{(1/2)}$ and $\hat{U}_P^{(1/2)}$ are respectively identified with $Z$ and $C$, the $\mathbb{Z}_2$-even modes have two-dimensional irreducible representation of $D_4\simeq (\mathbb{Z}_2\times \mathbb{Z}_2^{(\hat{U}_{Y})}) \rtimes \mathbb{Z}^{(\hat{U}_{P})}_2$, where $\mathbb{Z}_2$ acts on matter fields as diag$(-1,-1)$.

The mass matrix of right-handed neutrinos is of the form:
\begin{align}
    m^{(2-2-4)} = c_{(2-2-4)}^2
    \begin{pmatrix}
        X_3 & 0 & X_1\\
        0 & -\sqrt{2}X_2 & 0\\
        X_1 & 0 & X_3
    \end{pmatrix}
    ,
    \label{eq:MN_224}
\end{align}
where 
\begin{align}
    & X_1 =(\eta^{(16)}_0+\eta^{(16)}_8)^2+(\eta^{(16)}_4+\eta^{(16)}_{12})^2,
    \nonumber \\
    &X_2=\frac{1}{\sqrt{2}}\left((\eta^{(16)}_2+\eta^{(16)}_{10})+(\eta^{(16)}_6+\eta^{(16)}_{14})\right)^2,\\
    & X_3=2(\eta^{(16)}_0+\eta^{(16)}_8)(\eta^{(16)}_4+\eta^{(16)}_{12}),
    \nonumber
\end{align}
with
\begin{align}
    &\eta^{(n)}_N=\vartheta\left[\begin{array}{cc}
         \frac{N}{n} \\
         0
    \end{array}\right](0,n\tau).
\end{align}
We find that under these transformations \eqref{eq:M=4_even_trf} and \eqref{eq:M=2_even_trf}, the mass matrix of right-handed neutrino \eqref{eq:MN_224} is unchanged. 
Hence, there exists $D_4$ symmetry in the Majorana mass matrix. 
Note that $X_1$ would vanish by the selection rules due to the $\mathbb{Z}_4^{(Z)}$ and its $\mathbb{Z}_2$ gauging non-invertible symmetry, which appears in the zero modes with $M_N=4$ themselves at the tree level.
That is because $\psi^{0,4}$ and $\psi^{2,4}$ have the different $\mathbb{Z}_4^{(Z)}$ charges.
Such a selection rule due to $\mathbb{Z}_4^{(Z)}$ can be violated by charged zero modes, $\beta_i$ and $\gamma_i$ with $M_\beta=M_\gamma=2$.
Thus, the only $D_4$ symmetry remains due to D-brane instanton effects 
as well as the loop effects.

We have studied an example of D-brane instanton effects leading to right-handed neutrino mass terms.
There is another D-brane instanton configuration that possibly leads to 
right-handed neutrino mass terms, i.e., $3-3-6$ model.
This case is included in the generic discussions of Sec.~\ref{sec:n-point}, 
and the $D_4$ symmetry can be broken down.

\subsection{3-point couplings}
\label{sec:3-point}

As a next non-trivial example, let us discuss the 3-point coupling term with singlet fields on concrete magnetized D-brane models:
\begin{align}
e^{-S_{\mathrm{cl}}}\int d^3 \beta d^3\gamma e^{-\sum_{ija} d^{i,j,a}_{T^2/\mathbb{Z}_2}\beta_i\gamma_j N_a},
\label{eq:3-point}
\end{align}
where $\beta_i, \gamma_j$ are supposed to have three charged zero modes, which are induced by D-brane instanton, $N_a$ are singlet fields including right-handed neutrinos, and $d^{i,j,a}_{T^2/\mathbb{Z}_2}$ are their 3-point couplings.

To obtain 3-point couplings, we require three charged $\mathbb{Z}_2$-even or -odd zero modes for $\beta$ and $\gamma$. 
A minimal choice to realize the 3-point coupling of $N_a$ is $M_N=8$ for the flux of singlet field $N_a$ under which the magnetic flux of $\beta$ and $\gamma$ is $M_\beta=M_\gamma=4$.

\paragraph{4-4-8 model}\,\\

Let us examine the case, where the singlet fields corresponding to the $\mathbb{Z}_2$-even modes have $M_N=8$ and vanishing SS phase.
Then, the charged zero modes must have $M_{\beta}=4$ and $M_{\gamma}=4$.
Only if the number of the charged zero modes for each of $\beta_i$ and $\gamma_j$ is equal to three, the 3-point couplings can be induced.
We find the unique possibility that the charged zero modes are $\mathbb{Z}_2$ even and have vanishing SS phases. 
Three $\mathbb{Z}_2$-even modes for $\beta_i$ and $\gamma_j$ with $i,j=0,1,2$. 
The transformations of the $\mathbb{Z}_2$-even modes under $\hat{U}_Y^{(1/2)}$ and $\hat{U}_P^{(1/2)}$ are given in Eq.~\eqref{eq:M=4_even_trf} for $\beta_{i}, \gamma_{i}$ with $i,j=0,1,2$, and 
\begin{align}
    \hat{U}_Y^{(1/2)}\,&:\,        
    |\Psi\rangle_{\mathbb{Z}_2, +}^{M=8}
    \rightarrow 
        \begin{pmatrix}
        1 & 0 & 0 & 0 & 0\\
        0 & -1 & 0 & 0 & 0\\
        0 & 0 & 1 & 0 & 0\\
        0 & 0 & 0 & -1 & 0\\
        0 & 0 & 0 & 0 & 1\\
    \end{pmatrix}
    |\Psi\rangle_{\mathbb{Z}_2, +}^{M=8}
    \,,
    \nonumber\\
   \hat{U}_P^{(1/2)}\,&:\,       
    |\Psi\rangle_{\mathbb{Z}_2, +}^{M=8}
    \rightarrow 
        \begin{pmatrix}
        0 & 0 & 0 & 0 & 1\\
        0 & 0 & 0 & 1 & 0\\
        0 & 0 & 1 & 0 & 0\\
        0 & 1 & 0 & 0 & 0\\
        1 & 0 & 0 & 0 & 0\\
    \end{pmatrix}
    |\Psi\rangle_{\mathbb{Z}_2, +}^{M=8}
    \,,
\label{eq:M=8_even_trf}
\end{align}
for $N_a$ with $a=0,1,2,3,4$. 

After integrating out the charged zero modes, we obtain the 3-point coupling term:
\begin{align}
    e^{-S_{\mathrm{cl}}}\sum_{a,b,c}f_{a,b,c}N_{a}N_{b}N_{c},
\end{align}
with 
\begin{align}
     f_{a,b,c}\equiv -\left(\frac{1}{6}\right)^2\sum_{i_1,i_2,i_3,j_1,j_2,j_3}\epsilon_{i_1i_2i_3}\epsilon_{j_1j_2j_3}d_{T^2/\mathbb{Z}_2}^{i_1,j_1,a}d_{T^2/\mathbb{Z}_2}^{i_2,j_2,b}d_{T^2/\mathbb{Z}_2}^{i_3,j_3,c}.
\end{align}
In the $4-4-8$ model, 
the non-vanishing couplings are given by
\begin{align}
   & f_{0,0,0},\quad f_{0,0,2},\quad f_{0,0,4},\quad f_{0,1,1},\quad
    f_{0,1,3},\quad f_{0,2,2},\quad f_{0,2,4},\quad
    f_{0,3,3},\quad f_{0,4,4},\nonumber\\
   & f_{1,1,2},\quad f_{1,1,4},\quad
    f_{1,2,3},\quad f_{1,3,4},\quad f_{2,2,2},\quad f_{2,2,4},\quad f_{2,4,4},\quad
    f_{3,2,3},\quad f_{3,3,4},\nonumber\\
    &f_{4,4,4}.
\label{eq:4-4-8}
\end{align}
The explicit forms are written in Appendix \ref{app}. 
Let us fist consider $\hat{U}^{(1/2)}_Y$ transformation of these 3-point couplings. According to Eq. \eqref{eq:4-4-8}, all of the nonvanishing couplings satisfy that $a+b+c$ is even, so we find in the $4-4-8$ model that the 3-point coupling term is invariant under the $\hat{U}^{(1/2)}_Y$ transformation. 
Next, we study the $\hat{U}^{(1/2)}_P$ transformation under which the 3-point coupling transforms as
\begin{align}
f_{a,b,c}\xrightarrow{\hat{U}^{(1/2)}_P}f_{4-a,4-b,4-c}.
\end{align}
From the explicit form of 3-point couplings shown in Appendix \ref{app}, we find that they satisfy $f_{4-a,4-b,4-c}=f_{a,b,c}$ which results in the invariance of 3-point coupling under the $\hat{U}^{(1/2)}_P$ transformation.

We have studied an example of D-brane instanton effects inducing the 3-point coupling terms.
There is another D-brane instanton configuration possibly leading to 
3-point couplings, i.e., $5-5-10$ model.
This case is included in generic discussion of Sec.~\ref{sec:n-point}, 
and the $D_4$ symmetry can be broken down. 
Furthermore, this analysis can be generalized to $n$-point coupling in $N-N-2N$ model with arbitrary $N$, as seen in the next section.

\subsection{$n$-point couplings}
\label{sec:n-point}

In this section, we extend the analysis of Secs. \ref{sec:2-point} and \ref{sec:3-point} to generic $n$-point couplings and study whether the $D_4$ symmetry remains in the effective action. 
In particular, we examine the invariance of $n$-point couplings under the $D_4$ symmetry with an emphasis on the $N-N-2N$ model. 

Generic $n$-point couplings of singlet fields are derived from
\begin{align}
e^{-S_{\mathrm{cl}}}\int d^n \beta d^n\gamma e^{-\sum_{ija} d^{i,j,a}_{T^2/\mathbb{Z}_2}\beta_i\gamma_j N_a},
\label{eq:n-point}
\end{align}
where $\beta_i, \gamma_j$ are supposed to have $n$ number of charged zero modes, which are induced by D-brane instanton, $N_a$ are singlet fields under the SM gauge group including right-handed neutrinos, and $d^{i,j,a}_{T^2/\mathbb{Z}_2}$ are their 3-point couplings of $\mathbb{Z}_2$-even or odd modes. We focus on $\mathbb{Z}_2$-even modes with trivial SS phases, $\alpha_{1,\tau}^{(a)}=0$ (mod 1)  for $N_a$ as mentioned above.
For the charged zero modes, we focus $\mathbb{Z}_2$-even modes with all the possible SS phases.\footnote{For the charged zero modes, $\mathbb{Z}_2$-odd modes may appear.
However, one can arrive at the same conclusions when we take the $\mathbb{Z}_2$-odd modes for the charged zero modes.} 
The SS phases must satisfy the constraint $\alpha_{1,\tau}^{(i)}+\alpha_{1,\tau}^{(j)}=\alpha_{1,\tau}^{(a)}$.
Note that we set the trivial SS phase for $\alpha_{1,\tau}^{(a)}=0$ (mod 1).
On top of that, the SS phases of $\beta_i$ and $\gamma_j$ must be chosen such that they have the same number of massless modes.
Then, we find that the SS phases of $\beta_i$ and $\gamma_j$ must be the same, and they satisfy $\alpha_{1,\tau}^{(i)}+\alpha_{1,\tau}^{(j)}=2\alpha_{1,\tau}^{(i)}$. 
After integrating out the charged zero modes, we obtain
\begin{align}
    e^{-S_{\mathrm{cl}}}\sum_{a_1,...,a_n}f_{a_1,a_2,\ldots,a_n}N_{a_1}N_{a_2}\dots N_{a_n},
\label{eq:N-point_1}
\end{align}
with 
\begin{align}
    f_{a_1,a_2,\ldots,a_n} \equiv -\frac{1}{(N!)^2}\sum_{i_1,\ldots,i_n, j_1,\ldots, j_n}\epsilon_{i_1,i_2,\ldots ,i_n}\epsilon_{j_1,j_2,\ldots,j_n}d^{i_1,j_1,a_1}_{T^2/\mathbb{Z}_2}d^{i_2,j_2,a_2}_{T^2/\mathbb{Z}_2}\ldots d^{i_n,j_n,a_n}_{T^2/\mathbb{Z}_2}.
\label{eq:N-point_2}
\end{align}

Let us first analyze the transformations given by $\hat{U}_Y^{(1/2)}$ for $\beta_i, \gamma_j$ and $N_a$. The matrix of these transformations has diagonal elements that alternate between $1$ and $-1$ as in Eq.~\eqref{eq:UYUP_trf} when these zero-modes have even fluxes. 
Based on these considerations, let us perform a transformation for 3-point coupling appearing in the exponent \eqref{eq:n-point}:
\begin{align}
\sum_{i,j,a}d^{i,j,a}_{T^2/\mathbb{Z}_2}\beta_i\gamma_j N_a \xrightarrow{\hat{U}_Y^{(1/2)}} &\sum_{i,j,a,k,l,b}d^{i,j,a}_{T^2/\mathbb{Z}_2}\delta^i_k\delta^j_l\delta^a_b\beta_k\gamma_l N_b (-1)^{i+j+a-2(\alpha_1^{(i)}+\alpha_1^{(j)}+\alpha_1^{(a)})}
\nonumber\\
&=\sum_{k,l,b}\,\,d^{k,l,b}_{T^2/\mathbb{Z}_2}\beta_k\gamma_l N_b (-1)^{k+l+b},
\label{eq:dijaUY}
\end{align}
where we use $\alpha_1^{(i)}+\alpha_1^{(j)}=\alpha_1^{(a)}$ with $\alpha_1^{(i)} = \alpha_1^{(j)}$, and the range of $i,j,a$ is respectively given by $i=0,1,\dots i_{\rm max}$, $j=0,1\dots j_{\rm max}$, $a=0,1,\dots a_{\rm max}$.  
Note that $i_{\rm max}, j_{\rm max}, a_{\rm max}$ are given by $N_{\rm gen}-1$ where $N_{\rm gen}$ is the number of 
linearly-independent zero modes as shown in Table \ref{tab:zeromodes}. 
It turns out that the selection rule of 3-point coupling terms after the transformation is determined by the value of $k+l+b$. 
If $k+l+b$ is even, the coupling after the transformation $\hat{U}_Y^{(1/2)}$ is exactly the same as the former one. 
On the other hand, if $k+l+b$ is odd, the term \eqref{eq:dijaUY} becomes zero. 
Hence, we check the value of coupling that satisfies $k+l+b=$ even. 
The explicit form of 3-point coupling $d^{i,j,a}_{T^2/\mathbb{Z}_2}$ in the $N-N-2N$ model is defined in Eq.~\eqref{eq:dija_Z2_SS}. 
Note that we have focused on the even number of magnetic flux for $\beta_i$ and $\gamma$. 
On the other hand, the transformations of $\hat{U}_Y^{(1/2)}$ and $\hat{U}_P^{(1/2)}$ for $\beta_i$ and $\gamma_j$ with odd fluxes are given by Eq. \eqref{eq:Trf_generic} with $n_Y/M =n_P/M =1/2$. 
It is clear that the 3-point coupling is not invariant under the $\hat{U}_Y^{(1/2)}$ and $\hat{U}_P^{(1/2)}$ transformation. So, the 3-point coupling is invariant under the $\hat{U}_Y^{(1/2)}$ transformation only when $N=\mathrm{even}$.

To show the invariance of 3-point coupling term under $\hat{U}_Y^{(1/2)}$, we explore the selection rule of $d^{i,j,a}_{T^2/\mathbb{Z}_2}$. 
Since $d^{i,j,a}_{T^2/\mathbb{Z}_2}$ is given by the product of $\mathbb{Z}_2$-even modes, it obeys
\begin{align}
    d^{i,j,a}_{T^2/\mathbb{Z}_2} = d^{N-i,N-j,2N-a}_{T^2/\mathbb{Z}_2}.
\label{eq:dija_Z2_prop1}
\end{align}
From the expression of Eq. \eqref{eq:dija_SS}, the selection rules of $d^{i,j,a}_{T^2}$ in the $N-N-2N$ model is determined by the value of $i-j$ and $\delta_{i+j-a,2Nl-Nm}$. Here, the index $2Nl-Nm$ appearing in $\delta_{i+j-a,2Nl-Nm}$ is a multiple of $N$, indicating that the selection rules of $d^{i,j,a}_{T^2}$ is determined by $i+j-a\equiv 0$ (mod $N$). 
Note that $d^{i,j,a}_{T^2/\mathbb{Z}_2}$ can be expanded in terms of eight 3-point couplings $d^{i,j,a}_{T^2}$ as shown in Eq.~\eqref{eq:dija_Z2_SS}; thereby yielding the selection rules of  $d^{i,j,a}_{T^2/\mathbb{Z}_2}$ in the $N-N-2N$ model:
    \begin{align}
    i+\alpha_1^{(i)}+j+\alpha_1^{(j)}-a-\alpha_1^{(a)}&\equiv 0\qquad (\mathrm{mod}\,N)\qquad {\rm for}\,\,
        d^{i,j,a}_{T^2} ,\, d^{N-i,N-j,2N-a}_{T^2} \subset d^{i,j,a}_{T^2/\mathbb{Z}_2},
        \nonumber\\
        i+\alpha_1^{(i)}+j+\alpha_1^{(j)}+a+\alpha_1^{(a)}&\equiv 0\qquad (\mathrm{mod}\,N)\qquad {\rm for}\,\,d^{N-i,j,a}_{T^2},\, d^{i,N-j,2N-a}_{T^2} \subset d^{i,j,a}_{T^2/\mathbb{Z}_2},
        \nonumber\\
        i+\alpha_1^{(i)}-j-\alpha_1^{(j)}-a-\alpha_1^{(a)}&\equiv 0\qquad (\mathrm{mod}\,N)\qquad {\rm for}\,\,d^{i,N-j,a}_{T^2},\,d^{N-i,j,2N-a}_{T^2} \subset d^{i,j,a}_{T^2/\mathbb{Z}_2},
        \nonumber\\
        i+\alpha_1^{(i)}-j-\alpha_1^{(j)}+a+\alpha_1^{(a)}&\equiv 0\qquad (\mathrm{mod}\,N)\qquad {\rm for}\,\,d^{i,j,2N-a}_{T^2},\,d^{N-i,N-j,a}_{T^2} \subset d^{i,j,a}_{T^2/\mathbb{Z}_2}.
\label{eq:ija}    
\end{align}
i.e.,
    \begin{align}
    i+j+a =
    \left\{
    \begin{array}{lll}
         2a  &(\mathrm{mod}\,N)& {\rm for}\,\,
        d^{i,j,a}_{T^2} ,\, d^{N-i,N-j,2N-a}_{T^2} \subset d^{i,j,a}_{T^2/\mathbb{Z}_2}
        \\
        -2\alpha_1^{(a)}&(\mathrm{mod}\,N) &{\rm for}\,\,d^{N-i,j,a}_{T^2},\, d^{i,N-j,2N-a}_{T^2} \subset d^{i,j,a}_{T^2/\mathbb{Z}_2}
        \\
        2(j+a)+2\alpha_1^{(j)}&(\mathrm{mod}\,N) 
        &{\rm for}\,\,d^{i,N-j,a}_{T^2},\,d^{N-i,j,2N-a}_{T^2} \subset d^{i,j,a}_{T^2/\mathbb{Z}_2}
        \\
        2j -2\alpha_1^{(i)}&(\mathrm{mod}\,N)
        &{\rm for}\,\,d^{i,j,2N-a}_{T^2},\,d^{N-i,N-j,a}_{T^2} \subset d^{i,j,a}_{T^2/\mathbb{Z}_2}
    \end{array}
    \right.
    .
\end{align}
Here, we use $\alpha_1^{(i)}+\alpha_1^{(j)}=\alpha_1^{(a)}$ with $\alpha_1^{(i)} = \alpha_1^{(j)}$. 
It turns out that
\begin{align}
    d^{i,j,a}_{T^2/\mathbb{Z}_2}(-1)^{i+j+a} = 
    d^{i,j,a}_{T^2/\mathbb{Z}_2}\quad \mathrm{for}\quad N = {\rm even},\quad \alpha_1^{(i)}= \alpha_1^{(j)}\equiv 0\,(\mathrm{mod}\,1),\quad \alpha_1^{(a)}=2\alpha_1^{(i)}.
\end{align}
Hence, the 3-point coupling term \eqref{eq:dijaUY} is invariant under $\hat{U}_Y^{(1/2)}$ only when $N=$ even and trivial SS phases of matters.

We move to analyze the transformations of $\hat{U}_P^{(1/2)}$ in the $N-N-2N$ model. 
The numbers ($i,j,a$) run over $i=0,1,\dots,i_{\rm max}$, $j=0,1,\dots,j_{\rm max}$ and $a=0,1\dots,a_{\rm max}$. 
Then, the transformation $\hat{U}_P^{(1/2)}$ gives rise to
\begin{align}
\sum_{i,j,a}d^{i,j,a}_{T^2/\mathbb{Z}_2}\beta_i\gamma_j N_a 
\xrightarrow{\hat{U}_P^{(1/2)}}
\sum_{i,j,a,k,l,b}&\,d^{i,j,a}_{T^2/\mathbb{Z}_2}\delta_{i+k,i_{\rm max}}\delta_{j+l,j_{\rm max}}\delta_{a+b,a_{\rm max}}\beta_k\gamma_l N_b
\nonumber\\
=\sum_{k,l,b}&\,d^{k_{\rm max}-k, l_{\rm max}-l,b_{\rm max}-b}_{T^2/\mathbb{Z}_2}\beta_k\gamma_lN_b,
\label{eq:dijaYP}
\end{align}
with $k_{\rm max}=i_{\rm max}$, $l_{\rm max}= j_{\rm max}$ and $b_{\rm max}=a_{\rm max}$. 
To show the invariance of the 3-point coupling term under $\hat{U}_P^{(1/2)}$, 
the following property of 3-point coupling is required
\begin{align}
d^{k_{\rm max}-k, l_{\rm max}-l,b_{\rm max}-b}_{T^2/\mathbb{Z}_2}=
d^{k, l, b}_{T^2/\mathbb{Z}_2}.
\label{eq:dijaYP_prop1}
\end{align}
From the calculation of 3-point coupling in Appendix \ref{app:dijk}, we find that 
the 3-point coupling term is invariant under $\hat{U}_P^{(1/2)}$, only when $N=$ even and the 
SS phases of chiral matters are trivial.

So far, we have discussed the invariance of 3-point couplings under $\hat{U}_Y^{(1/2)}$ and $\hat{U}_P^{(1/2)}$ which will lead to the invariance of generic $n$-point couplings \eqref{eq:N-point_2} after integrating out zero modes $\beta$ and $\gamma$. 
In the following analysis, we verify whether the $n$-point coupling \eqref{eq:N-point_2} is invariant under $\hat{U}_Y^{(1/2)}$ and $\hat{U}_P^{(1/2)}$. 
To see this, we first consider transformation of $n$-point coupling under $\hat{U}_Y^{(1/2)}$:
\begin{align}
  &\sum_{a_1,\ldots,a_n}f^{a_1,a_2,\dots, a_n} N_{a_1}N_{a_2}\dots N_{a_n} 
  \nonumber\\
  \xrightarrow{\hat{U}_Y^{(1/2)}}&
    \sum_{a_1,\ldots,a_n, b_1, \ldots, b_n}f^{a_1,a_2,\dots ,a_n}\delta^{a_1}_{b_1}\delta^{a_2}_{b_2}\dots\delta^{a_n}_{b_n}(-1)^{a_1+\cdots +a_n}N_{b_1}N_{b_2}\dots N_{b_n}
    \nonumber\\
    &=\,\,\sum_{b_1, \ldots, b_n}f^{b_1,b_2,\dots,b_n}(-1)^{b_1+\dots+b_n}N_{b_1}N_{b_2}\dots N_{b_n},
\label{eq:N-point_UY}
\end{align}
where it is notable that the SS phases of $N_a$ are chosen to be zero. 
It indicates that if the $n$-point coupling is invariant under the transformation of $\hat{U}_Y^{(1/2)}$, the non-vanishing terms of $f^{b_1,b_2,\dots,b_n}$ should satisfy $\sum_{m=1}^n b_m=$ even. 
From Eq. \eqref{eq:ija}, $b_m$ obeys
\begin{align}
    b_m +\alpha_1^{(b_m)}=
    \left\{
    \begin{array}{ll}
         \pm (i_m + j_m + 2\alpha_1^{(i_m)})  & (\mathrm{mod}\, N)\\
         \pm(i_m - j_m )  & (\mathrm{mod}\, N)\\
    \end{array}
    \right.
,
\end{align}
which are useful to simplify the nonvanishing $n$-point coupling \eqref{eq:N-point_2}:
\begin{align}
    &-(N!)^2f_{b_1,b_2,\ldots,b_n} (-1)^{\sum_{m=1}^n b_m} 
    \nonumber\\
    &=\left\{
    \begin{array}{l}
       \sum_{i_1,\ldots,i_n, j_1,\ldots, j_n}\epsilon_{i_1,i_2,\ldots ,i_n}\epsilon_{j_1,j_2,\ldots,j_n}d^{i_1,j_1,b_1}_{T^2/\mathbb{Z}_2}\ldots d^{i_n,j_n,b_n}_{T^2/\mathbb{Z}_2} (-1)^{N}\\
\sum_{i_1,\ldots,i_n, j_1,\ldots, j_n}\epsilon_{i_1,i_2,\ldots ,i_n}\epsilon_{j_1,j_2,\ldots,j_n}d^{i_1,j_1,b_1}_{T^2/\mathbb{Z}_2}\ldots d^{i_n,j_n,b_n}_{T^2/\mathbb{Z}_2} (-1)^{N-\sum_{m=1}^n \alpha_1^{(b_m)}}
    \end{array}
\right.
.
\label{eq:f_prop}
\end{align}
Here, we use the fact that the anti-symmetric tensor is non zero when $\sum_{m=1}^n i_m = \sum_{m=1}^n j_m =n(n+1)/2$. 
Note that the nonvanishing $n$-point couplings should have the same structure for the first and second lines of Eq.~\eqref{eq:f_prop}. 
Hence, we find the following equality for the nonvanishing $n$-point coupling: 
\begin{align}
    f_{b_1,b_2,\ldots,b_n} (-1)^{\sum_{m=1}^n b_m} = 
    \left\{
\begin{array}{cc}
    f_{b_1,b_2,\ldots,b_n} & (N=\mathrm{even},\quad \sum_m \alpha^{(b_m)}_1 \equiv 0\;(\mathrm{mod}\;2))\\
    -f_{b_1,b_2,\ldots,b_n} & (N=\mathrm{odd},\quad \sum_m \alpha^{(b_m)}_1 \equiv 0\;(\mathrm{mod}\;2))
\end{array}
\right.
.
\label{eq:N-point_UY_1}
\end{align}
Thus, when $N=$ even, the $n$-point coupling term is invariant under the transformation of $\hat{U}^{(1/2)}_Y$. 
On the other hand, when $N=$ odd, the $n$-point coupling term is not invariant under $\hat{U}^{(1/2)}_Y$. 

Next, we analyze the transformation of $n$-point coupling under $\hat{U}_P^{(1/2)}$:
\begin{align}
  &\sum_{a_1,\ldots,a_n}f^{a_1,a_2,\dots, a_n} N_{a_1}N_{a_2}\dots N_{a_n} 
  \nonumber\\
  \xrightarrow{ \hat{U}_P^{(1/2)}}&
    \sum_{a_1,\ldots,a_n, b_1, \ldots, b_n}f^{a_1,a_2,\dots ,a_n}\delta_{b_1+a_1,(b_1)_{\rm max}}\delta_{b_2+a_2,(b_2)_{\rm max}}\dots\delta_{b_n+a_n,(b_n)_{\rm max}}N_{b_1}N_{b_2}\dots N_{b_n}
    \nonumber\\
    &=\,\,\sum_{b_1, \ldots, b_n}f^{(b_1)_{\rm max}-b_1,(b_2)_{\rm max}-b_2,\dots,(b_n)_{\rm max}-b_n}N_{b_1}N_{b_2}\dots N_{b_n},  
\label{eq:N-point_UP}
\end{align}
with $(a_m)_{\rm max}=(b_m)_{\rm max}$ for $m=1,2,...,n$. 
The $n$-point couplings are rewritten as
\begin{align}
    &f^{(b_1)_{\rm max}-b_1,(b_2)_{\rm max}-b_2,\dots,(b_n)_{\rm max}-b_n}
    \nonumber\\
    &=\sum_{i_1,\dots, i_n, l_1,\dots, l_n}\epsilon_{i_1,\dots, i_n}\epsilon_{j_1,\dots,j_n}\cdot d^{i_1,j_1,(b_1)_{\rm max}-b_1}\dots d^{i_n,j_n,(b_n)_{\rm max}-b_n}
    \nonumber\\
    &=\sum_{i_1,\dots, i_n, l_1,\dots, l_n}\epsilon_{(i_1)_{\rm max} -i_1,\dots ,(i_n)_{\rm max}-i_n}\epsilon_{(j_1)_{\rm max}-j_1,\dots,(j_n)_{\rm max}-j_n}
    \nonumber\\
    &\times d^{(i_1)_{\rm max}-i_1,(j_1)_{\rm max}-j_1,(b_1)_{\rm max}-b_1}\dots d^{(i_n)_{\rm max}-i_n,(j_n)_{\rm max}-j_n,(b_n)_{\rm max}-b_n}
    \nonumber\\
    &=\sum_{i_1,\dots, i_n, l_1,\dots, l_n}\epsilon_{(i_1)_{\rm max} -i_1,\dots ,(i_n)_{\rm max}-i_n}\epsilon_{(j_1)_{\rm max}-j_1,\dots,(j_n)_{\rm max}-j_n} d^{i_1,j_1,b_1}\dots d^{i_n,j_n,b_n},
\end{align}
where we use Eq.~\eqref{eq:dijaYP_prop1} in the third equality which holds for $N=$ even and trivial SS phases of matters. 
By using the property of the anti-symmetric tensor:
\begin{align}
    \epsilon_{(i_1)_{\rm max} -i_1,\dots ,(i_n)_{\rm max}-i_n} = (-1)^{n(n+1)/2}\epsilon_{i_1,i_2,\dots,i_n},
\end{align}
with $(i_m)_{\rm max}-i_m$ and $i_m$ $(m=1,2,\cdots n)$ being integers within $0$ and $n$, we find that the $n$-point coupling \eqref{eq:N-point_2} is indeed invariant under $\hat{U}_P^{(1/2)}$ for $N=$ even, i.e.,
\begin{align}
    f^{(b_1)_{\rm max}-b_1,(b_2)_{\rm max}-b_2,\dots,(b_n)_{\rm max}-b_n}
    &=f^{b_1,b_2,\dots,b_n}.
\end{align}
We conclude that $n$-point coupling terms in the $N-N-2N$ model enjoy the $D_4$ symmetry at the non-perturbative level for $N=$ even and trivial SS phases of chiral matters. 
On the other hand, no symmetry remains for $N=$ odd.

\subsection{$\mu$-term}
\label{sec:mu-term}

In this section, we move to the selection rule of $\mu$-terms induced by the D-brane instanton. 
The $\mu$-terms are the mass terms of Higgs fields, which 
have the $SU(2)$ doublet representations as well as $U(1)_Y$ charges in contrast to the analysis of Sec.~\ref{sec:2-point}.
The $\mu$-terms can be derived from the D-brane instanton effects:
\begin{align}
    e^{-S_{\mathrm{cl}}}\int\mathcal{D}\alpha\mathcal{D}\gamma\mathcal{D}\beta e^{y_a^u\alpha\cdot H_u^a\beta+y_b^d\alpha\cdot H_d^b\gamma}=e^{-S_{\mathrm{cl}}}y_a^uy_b^dH_u^aH_d^b.
\end{align}
We assume two pairs of Higgs fields, i.e., $a=1,2$. 
The pair of charged zero modes is restricted to one, otherwise the $\mu$-term cannot be obtained. 
For instance, when we choose the magnetic flux of Higgs fields as four, it is possible to consider other cases: i) $M_\alpha=M_\beta=M_\gamma=2$, ii) $M_\alpha=1$, $M_\beta=M_\gamma=3$, and iii) $M_\alpha=3$, $M_\beta=M_\gamma=1$. However in these cases, one cannot generate the $\mu$-term. 
The coupling $y_a^u$ is the Yukawa coupling of, $\alpha$, $H_u^a$ and $\beta$, while $y_b^d$ is that of $\alpha$, $H_d^b$ and $\gamma$. 
Since the Yukawa coupling is written as
\begin{align}
    y^u_a=d^{00a}_{T^2/\mathbb{Z}_2},
\end{align}
the $\mu$-term is given by when the SS phases are $(\alpha^{(0)}_1,\alpha^{(0)}_\tau)=(0,0),(0,1/2),(1/2,0)$:
\begin{align}
   \nonumber \mu_{ab}&=e^{-S_{\mathrm{cl}}}d^{00a}_{T^2/\mathbb{Z}_2}d^{00b}_{T^2/\mathbb{Z}_2}\\
  \nonumber &=e^{-S_{\mathrm{cl}}}c_{(1-1-2)}^2\mathrm{exp}\left\{2\pi i\left(2\alpha^{(0)}_
  1\alpha^{(0)}_\tau+2\alpha^{(0)}_1\alpha^{(0)}_\tau-\frac{(a+\alpha^{(a)}_1)\alpha^{(a)}_\tau+(b+\alpha^{(b)}_1)\alpha^{(b)}_\tau}{2}\right)\right\}\\
&\times\sum^1_{m=0}\vartheta\left[\begin{array}{cc}
        \frac{m}{2}  \\
        0
   \end{array}\right]
   (0,2\tau)\delta_{-a,2l-m}
   \times \sum^1_{k=0}\vartheta\left[\begin{array}{cc}
         \frac{k}{2}  \\
        0
   \end{array}\right]
   (0,2\tau)\delta_{-b,2l-k},
\end{align}
which depends on the SS phase $\alpha^{(0)}_1$ and $\alpha^{(1)}_1$. 
For the other case with the SS phase $(\alpha^{(0)}_1,\alpha^{(0)}_\tau)=(1/2,1/2)$, $\mu$-term matrix $\mu_{ab}$ vanishes because the 3-point coupling becomes zero. 
It was known in Ref. \cite{Hoshiya:2021nux} that when the number of Higgs pairs is two, there are four possibilities for the SS phases: $(\alpha_1^{(0)},\alpha_\tau^{(0)})=(0,0),(0,1/2),(1/2,0),(1/2,1/2)$. Note that $\alpha,\beta,\gamma$ are even mode when $(\alpha_1^{(0)},\alpha_\tau^{(0)})=(0,0),(1/2,0)$ and odd mode when $(\alpha_1^{(0)},\alpha_\tau^{(0)})=(0,1/2),(1/2,1/2)$. The Higgs field has two even-modes when $(\alpha_1^{(0)},\alpha_\tau^{(0)})=(0,0)$ and has an even-mode and an odd-mode when $(\alpha_1^{(0)},\alpha_\tau^{(0)})=(0,1/2),(1/2,0),(1/2,1/2)$.
The normalization factor of each wavefunction is $1/2$ in all of the cases, and it is irrelevant to the following discussions. 
In each case, we derive the $\mu$-term matrix in the following.

First, we discuss the case with $(\alpha_1^{(0)},\alpha_\tau^{(0)})=(0,0)$. 
Then, the $\mu$-term matrix is of the form:
\begin{align}
    \mu^{(0,0)}=e^{-S_{\mathrm{cl}}}c_{(1-1-2)}^2\begin{pmatrix}
        Z_1&Z_3\\
        Z_3&Z_2\\
    \end{pmatrix}
\label{eq:mu_ex1}
    ,
\end{align}
where $Z_1,Z_2,Z_3$ is 
\begin{align}
   Z_1 &=\vartheta\left[\begin{array}{cc}
         0 \\
         0 
    \end{array}\right](0,2\tau)\times
    \vartheta\left[\begin{array}{cc}
        0  \\
         0 
    \end{array}\right](0,2\tau),
    \nonumber\\
    Z_2&=\vartheta\left[\begin{array}{cc}
         1/2 \\
         0 
    \end{array}\right](0,2\tau)\times
    \vartheta\left[\begin{array}{cc}
        1/2  \\
         0 
    \end{array}\right](0,2\tau),
    \nonumber\\
    Z_3&=\vartheta\left[\begin{array}{cc}
         1/2 \\
         0 
    \end{array}\right](0,2\tau)\times
    \vartheta\left[\begin{array}{cc}
         0  \\
         0 
    \end{array}\right](0,2\tau).
\end{align}
Note that one can obtain the same $\mu$-term matrix for the $(\alpha_1^{(0)},\alpha_\tau^{(0)})=(1/2,0)$ case. 
When we consider the transformation of $\hat{U}^{(1/2)}_Y$ and $\hat{U}^{(1/2)}_P$: 
\begin{align}
   & \hat{U}^{(1/2)}_Y\: : \: \mu \xrightarrow{\hat{U}^{(1/2)}_Y}\mu^\prime=\begin{pmatrix}
        Z_1&-Z_3\\
        -Z_3&Z_2\\
    \end{pmatrix}
    ,\\
   & \hat{U}^{(1/2)}_P\: : \: \mu\xrightarrow{\hat{U}^{(1/2)}_P}\mu^\prime=\begin{pmatrix}
       Z_2&Z_3\\
       Z_3&Z_1\\
   \end{pmatrix}
   ,
\end{align}
the $\mu$-term matrix is not invariant under $\hat{U}^{(1/2)}_Y$ and $\hat{U}^{(1/2)}_P$.

Next, we consider the case with $(\alpha_1^{(0)},\alpha_\tau^{(0)})=(0,1/2)$. 
The $\mu$-term matrix is given by
\begin{align}
    \mu_{ab}^{(0,1/2)}=e^{-\pi i(a+b)} \mu_{ab}^{(0,0)}= e^{-S_{\mathrm{cl}}}c^2_{(1-1-2)}\begin{pmatrix}
        Z_1&-Z_3\\
        -Z_3&Z_2\\
    \end{pmatrix}
    ,
\end{align}
which corresponds to Eq.~\eqref{eq:mu_ex1} with swapping $Z_1$ and $Z_2$. We also find that the $\mu$-term matrix vanishes when $(\alpha_1^{(0)},\alpha_\tau^{(0)})=(1/2,1/2)$. 
Hence, the $\mu$-term matrix is not still invariant under $\hat{U}^{(1/2)}_Y$ or $\hat{U}^{(1/2)}_P$.\footnote{If all of the instanton effects could appear in the model and we would sum all the configurations of D-brane instantons inducing the charged zero modes with proper weights, the total $\mu$-term matrix would be given by $\mu_{\rm tot}\equiv \mu^{(0,0)} + \mu^{(1/2,0)} + 2\mu^{(0,1/2)}$, which is invariant under $\hat{U}^{(1/2)}_Y$. In addition, there is the $\mathbb{Z}_2$ symmetry such as $H^a_u \to - H^a_u$ and $H^a_d \to - H^a_d$. Thus, the $D_4$ symmetry could be broken to $\mathbb{Z}_2\times \mathbb{Z}_2^{(Z)}$ symmetry by D-brane instanton effects.}

The result of the Higgs $\mu$-term matrix is different from the remaining symmetry of the singlets $N_a$ with $M_N=4$.
This difference may originate from the fact that the Higgs fields are doublets while $N_a$ are singlets and the number of charged zero modes are different from each other.



\section{Conclusions and discussions}
\label{sec:con}

It was shown in Ref. \cite{Kobayashi:2024yqq} that the non-invertible symmetry associated with discrete isometry operators appears in the low-energy effective action after gauging the orbifold group in toroidal orbifold compactifications of type IIB magnetized D-brane models. 
When the chiral matters feel an even number of magnetic flux on $T^2/\mathbb{Z}_2$, they belong to the representations of $D_4$ flavor symmetry of the non-invertible symmetry. 

In this paper, we extended the IIB setup \cite{Kobayashi:2024yqq} to its T-dual intersecting D-brane models. 
We found the same operators inducing the non-invertible symmetry which controls the flavor symmetry of twisted fields. 
However, it was proposed in Refs. \cite{Heckman:2024obe,Kaidi:2024wio} that the non-invertible symmetry at tree level will be broken by quantum effects. 
It motivates us to incorporate quantum corrections in the type IIA intersecting and type IIB magnetized D-brane models. 
Since the chiral matters are labeled by a conjugacy class of the gauged orbifold group, their selection rules are expected to be different from the tree-level ones. 

Following the selection rules of conjugacy class discussed in Sec. \ref{sec:generic-loop}, we proposed the selection rules of chiral matters at loop level in Sec. \ref{sec:IIB_loop}. 
It turned out that the $D_4$ flavor symmetry still remains even when we include radiative corrections, but the other symmetries are broken down. 
Then, it is of particular interest whether the $D_4$ symmetry exists at the non-perturbative level. 

In Sec. \ref{sec:np}, we analyzed D-brane instanton effects in the context of type IIB magnetized D-brane models on $T^2/\mathbb{Z}_2$. 
In particular, we focused on the $n$-point couplings of SM singlet fields in Sec. \ref{sec:n-point} and $\mu$-term in Sec. \ref{sec:mu-term}. 
We studied possible configurations of D-brane instantons systematically.
For the SM singlet fields $N_a$, it was argued in Sec. \ref{sec:IIB_loop} that there exists the $D_4$ symmetry in the effective action at loop level when $M_N=2\,\mathbb{Z}$, but the analysis of Sec. \ref{sec:n-point} shows that the $n$-point couplings induced by D-brane instantons do not enjoy the $D_4$ symmetry in a generic value of magnetic flux. 
On the other hand, there still exists the $D_4$ symmetry when $M_{N}=4\,\mathbb{Z}$. Note that when $M_{N}=4\,\mathbb{Z}$, all of the zero modes correspond to the $D_4$ singlets, and doublets are not included.
The flavor symmetry of these singlets is the Abelian $\mathbb{Z}_2\times \mathbb{Z}_2$ symmetry. For the $\mu$-term, the $D_4$ symmetry could be broken to $\mathbb{Z}_2\times \mathbb{Z}_2$ symmetry.

Here, we discuss our results from the viewpoint of anomaly.
In a generic gauge theory, non-Abelian discrete symmetries as well as Abelian discrete symmetries can be anomalous and broken by gauge instanton effects \cite{Krauss:1988zc,Ibanez:1991hv,Banks:1991xj,Araki:2008ek,Chen:2015aba,Kobayashi:2021xfs}.
Following Ref.~\cite{Araki:2008ek}, we can examine anomalies in our setup.
However, the $D_4$ symmetry is free from the gauge-gauge-$D_4$ mixed anomalies.
The $D_4$ breaking due to D-brane instanton effects is caused by charged zero modes.
Thus, D-brane instanton effects may include more aspects than gauge instanton effects.
Note that D-brane instantons can not generate correction terms on the $T^2$ compactifications, because charged zero modes have a large supersymmetry.
In this sense, the $\mathbb{Z}_2$ gauging is important to induce correction terms, which violate the perturbative non-invertible symmetries to smaller ones. 
Alternatively, some D-brane instanton configurations may be forbidden in 
explicit models, although we have shown possible configurations.
It is important to study explicitly full models, but that is beyond our scope.

\acknowledgments

We thank S. Uemura for useful discussions in the early stage of this work. 
This work was supported in part by JSPS KAKENHI Grant Numbers JP23H04512 (H.O) and JP23K03375 (T.K.).

\appendix

\section{Explicit form of 3-point couplings in the 4-4-8 model}
\label{app}

We list the nonvanishing 3-point couplings in the 4-4-8 model:
\begin{align}
    f_{000}&=f_{444}=d_{T^2/\mathbb{Z}_2}^{000}d_{T^2/\mathbb{Z}_2}^{110}d_{T^2/\mathbb{Z}_2}^{220}=2^8d_{T^2}^{000}d_{T^2}^{130}d_{T^2}^{220},\nonumber\\
    f_{002}&=f_{442}=d_{T^2/\mathbb{Z}_2}^{000}d_{T^2/\mathbb{Z}_2}^{220}d_{T^2/\mathbb{Z}_2}^{112}=2^7d_{T^2}^{000}d_{T^2}^{220}d_{T^2}^{112},\nonumber\\
    f_{004}&=f_{440}=d_{T^2/\mathbb{Z}_2}^{000}d_{T^2/\mathbb{Z}_2}^{110}d_{T^2/\mathbb{Z}_2}^{224}+d_{T^2/\mathbb{Z}_2}^{000}d_{T^2/\mathbb{Z}_2}^{114}d_{T^2/\mathbb{Z}_2}^{220}+d_{T^2/\mathbb{Z}_2}^{004}d_{T^2/\mathbb{Z}_2}^{110}d_{T^2/\mathbb{Z}_2}^{220},\nonumber\\
    &=2^8(d_{T^2}^{000})^2d_{T^2}^{130}+2^8d_{T^2}^{000}d_{T^2}^{130}d_{T^2}^{220}+2^8(d_{T^2}^{220})^2d_{T^2}^{130},\nonumber\\
    f_{011}&=f_{433}=d_{T^2/\mathbb{Z}_2}^{220}d_{T^2/\mathbb{Z}_2}^{011}d_{T^2/\mathbb{Z}_2}^{101}+d_{T^2/\mathbb{Z}_2}^{000}d_{T^2/\mathbb{Z}_2}^{121}d_{T^2/\mathbb{Z}_2}^{211}\nonumber\\
    &=2^5d_{T^2}^{220}(d_{T^2}^{037}+d_{T^2}^{011})^2+2^5d_{T^2}^{000}(d_{T^2}^{231}+d_{T^2}^{217})^2\nonumber\\
    f_{022}&=f_{422}=d_{T^2/\mathbb{Z}_2}^{110}d_{T^2/\mathbb{Z}_2}^{202}d_{T^2/\mathbb{Z}_2}^{022}=2^6d_{T^2}^{130}(d_{T^2}^{202})^2,\nonumber\\
    f_{024}&=d_{T^2/\mathbb{Z}_2}^{000}d_{T^2/\mathbb{Z}_2}^{112}d_{T^2/\mathbb{Z}_2}^{224}+d_{T^2/\mathbb{Z}_2}^{220}d_{T^2/\mathbb{Z}_2}^{112}d_{T^2/\mathbb{Z}_2}^{004}\nonumber\\
    &=2^7(d_{T^2}^{000})^2d_{T^2}^{112}+2^7(d_{T^2}^{220})^2d_{T^2}^{112},\nonumber\\
    f_{033}&=f_{411}=d_{T^2/\mathbb{Z}_2}^{220}d_{T^2/\mathbb{Z}_2}^{013}d_{T^2/\mathbb{Z}_2}^{103}+d_{T^2/\mathbb{Z}_2}^{000}d_{T^2/\mathbb{Z}_2}^{123}d_{T^2/\mathbb{Z}_2}^{213}\nonumber\\
    &=2^5d_{T^2}^{220}(d_{T^2}^{217}+d_{T^2}^{231})^2+2^5d_{T^2}^{000}(d_{T^2}^{037}+d_{T^2}^{011})^2,\nonumber\\
    f_{112}&=f_{332}=d_{T^2/\mathbb{Z}_2}^{011}d_{T^2/\mathbb{Z}_2}^{121}d_{T^2/\mathbb{Z}_2}^{202}=2^4(d_{T^2}^{037}+d_{T^2}^{011})(d_{T^2}^{321}+d_{T^2}^{217})d_{T^2}^{202},\nonumber\\
    f_{123}&=d_{T^2/\mathbb{Z}_2}^{013}d_{T^2/\mathbb{Z}_2}^{202}d_{T^2/\mathbb{Z}_2}^{121}+d_{T^2/\mathbb{Z}_2}^{011}d_{T^2/\mathbb{Z}_2}^{123}d_{T^2/\mathbb{Z}_2}^{202}\nonumber\\
    &=2^4(d_{T^2}^{217}+d_{T^2}^{231})d_{T^2}^{202}(d_{T^2}^{231}+d_{T^2}^{217})+2^4(d_{T^2}^{037}+d_{T^2}^{011})d_{T^2}^{202}(d_{T^2}^{037}+d_{T^2}^{011}),\nonumber\\
    f_{222}&=d_{T^2/\mathbb{Z}_2}^{202}d_{T^2/\mathbb{Z}_2}^{022}d_{T^2/\mathbb{Z}_2}^{112}=2^5d_{T^2}^{112}(d_{T^2}^{202})^2,\nonumber\\
    &=2^5(d_{T^2}^{217}+d_{T^2}^{231})^2d_{T^2}^{000}+2^5(d_{T^2}^{037}+d_{T^2}^{011})^2d_{T^2}^{220},\nonumber\\
    f_{013}&=f_{431}=d_{T^2/\mathbb{Z}_2}^{220}d_{T^2/\mathbb{Z}_2}^{101}d_{T^2/\mathbb{Z}_2}^{013}+d_{T^2/\mathbb{Z}_2}^{000}d_{T^2/\mathbb{Z}_2}^{121}d_{T^2/\mathbb{Z}_2}^{213}\nonumber\\
    &=2^5d_{T^2}^{220}(d_{T^2}^{037}+d_{T^2}^{011})(d_{T^2}^{033}+d_{T^2}^{015})+2^5d_{T^2}^{000}(d_{T^2}^{231}+d_{T^2}^{217})(d_{T^2}^{037}+d_{T^2}^{235}).
    \end{align}

\section{Transformation of 3-point couplings under $\hat{U}^{(1/2)}_P$}
\label{app:dijk}

We consider $\hat{U}^{(1/2)}_P$ transformation of the 3-point coupling $d_{T^2/\mathbb{Z}_2}^{i,j,a}$ in the $N-N-2N$ model. According to Table 2, there are 8 cases:
\begin{enumerate}
    \item $N=\mathrm{even}$ and $(\alpha_1^{(i)},\alpha_\tau^{(i)})=(0,0)$ (mod 1)

    In this case, $i_{\mathrm{max}}$ and $a_{\mathrm{max}}$ are satisfied  $i_{\mathrm{max}}=N/2$ and $a_{\mathrm{max}}=N$. These have a relation: $2i_{\mathrm{max}}=a_{\mathrm{max}}$. Hence, the 3-point coupling $d^{ija}_{T^2/\mathbb{Z}_2}$ is invariant under $\hat{U}^{(1/2)}_P$ transformation, i.e.,
\begin{align}
    d^{i,j,a}_{T^2/\mathbb{Z}_2}\xrightarrow{\hat{U}^{(1/2)}_P}&\,d^{i_{\mathrm{max}}-i,j_{\mathrm{max}}-j,a_{\mathrm{max}}-a}_{T^2/\mathbb{Z}_2}\nonumber\\
    =&\,d^{\frac{N}{2}-i,\frac{N}{2}-j,N-a}_{T^2/\mathbb{Z}_2}\nonumber\\
    =&\,d^{\frac{N}{2}-i,\frac{N}{2}-j,N-a}_{T^2}+d^{\frac{N}{2}-i,\frac{N}{2}-j,N+a}_{T^2}+d^{\frac{N}{2}-i,\frac{N}{2}+j,N-a}_{T^2}+d^{\frac{N}{2}-i,\frac{N}{2}+j,N+a}_{T^2}\nonumber\\
    &\,+d^{\frac{N}{2}+i,\frac{N}{2}-j,N-a}_{T^2}+d^{\frac{N}{2}+i,\frac{N}{2}-j,N+a}_{T^2}+d^{\frac{N}{2}+i,\frac{N}{2}+j,N-a}_{T^2}+d^{\frac{N}{2}+i,\frac{N}{2}+j,N+a}_{T^2}\nonumber\\
    =&\,d^{-i,-j,-a}_{T^2}+d^{-i,-j,a}_{T^2}+d^{-i,j,-a}_{T^2}+d^{-i,j,a}_{T^2}\nonumber\\
    &\,+d^{i,-j,-a}_{T^2}+d^{i,-j,a}_{T^2}+d^{i,j,-a}_{T^2}+d^{i,j,a}_{T^2}\nonumber\\
    =&\,d^{i,j,a}_{T^2/\mathbb{Z}_2},
\end{align}
where we use $d^{i,j,a}_{T^2}=d^{i+h,j+h,a+2h}_{T^2}$ with $h$ being an arbitrary integer. 
Note that we suppose that the SS phases of $\beta_i$ and $\gamma_j$ are the same. This equality originates from the selection rule $d^{i,j,a}_{T^2}$, i.e., $i+j-a\equiv 0$ (mod $N$) in the $N-N-2N$ model.

    \item $N=\mathrm{even}$ and $(\alpha_1^{(i)},\alpha_\tau^{(i)})=(1/2,0)$ (mod 1)
    
    In the cases 3 and 4, the index of chiral zero modes runs from $0$ to 
\begin{align}
    i_{\mathrm{max}}=\left\{\begin{array}{cc}
         \frac{N}{2}-1,\quad\mathrm{for}\quad N=\mathrm{even}\\
         \frac{N-1}{2},\quad \mathrm{for}\quad N=\mathrm{odd}\\
    \end{array}\right.
    .
\end{align}
They do not satisfy $i_{\mathrm{max}}=2a_{\mathrm{max}}$, and we also check the transformation of 3-point coupling for $N=\mathrm{even}$:
\begin{align}
    d^{i,j,a}_{T^2/\mathrm{Z}_2}\xrightarrow{\hat{U}^{(1/2)}_P}&\,d^{i_{\mathrm{max}}-i,j_{\mathrm{max}}-j,a_{\mathrm{max}}-a}_{T^2/\mathbb{Z}_2}\nonumber\\
    =&\,d^{\frac{N}{2}-1-i,\frac{N}{2}-1-j,N-a}_{T^2/\mathbb{Z}_2}\nonumber\\
    =&\,d^{\frac{N}{2}-1-i,\frac{N}{2}-1-j,N-a}_{T^2}+d^{\frac{N}{2}-1-i,\frac{N}{2}-1-j,N+a}_{T^2}+d^{\frac{N}{2}-1-i,\frac{N}{2}+1+j,N-a}_{T^2}\nonumber\\
    &\,+d^{\frac{N}{2}-1-i,\frac{N}{2}+1+j,N+a}_{T^2}+d^{\frac{N}{2}+1+i,\frac{N}{2}-1-j,N-a}_{T^2}+d^{\frac{N}{2}+1+i,\frac{N}{2}-1-j,N+a}_{T^2}\nonumber\\
    &\,+d^{\frac{N}{2}+1+i,\frac{N}{2}+1+j,N-a}_{T^2}+d^{\frac{N}{2};1+i,\frac{N}{2}+1+j,N+a}_{T^2}\nonumber\\
    =&\,d^{-i,-j,2-a}_{T^2}+d^{-i,-j,2+a}_{T^2}d^{-i,2+j,2-a}_{T^2}+d^{-i,2+j,2+a}_{T^2}\nonumber\\
    &\,+d^{2+i,-j,2-a}_{T^2}+d^{2+i,-j,2+a}_{T^2}+d^{i,j,-2-a}_{T^2}+d^{i,j,-2+a}_{T^2}\nonumber\\
    \neq&\, d^{i,j,a}_{T^2/\mathbb{Z}_2}.
\end{align}
but it is not invariant under $\hat{U}^{(1/2)}_P$.

    \item $N=\mathrm{even}$ and $(\alpha_1^{(i)},\alpha_\tau^{(i)})=(0,1/2)$ (mod 1)

Third, we consider the $(\alpha^{(i)}_i,\alpha^{(i)}_\tau)=(0,1/2)$ case in which $i_{\mathrm{max}}$ takes following values:
\begin{align}
     i_{\mathrm{max}}=\left\{\begin{array}{cc}
         \frac{N}{2}-1,\quad\mathrm{for}\quad N=\mathrm{even}\\
         \frac{N-1}{2},\quad \mathrm{for}\quad N=\mathrm{odd}\\
    \end{array}\right.
    .
\end{align}
In this cases, we should consider $u^{(i,j,a)}$ factor as in $\eqref{eq:dija_Z2_SS}$, but the transformation law of $d^{i,j,a}_{T^2}$ is the same as 3 and 4, hence the 3-point coupling is not invariant under $\hat{U}_P^{(1/2)}$.

    \item $N=\mathrm{even}$ and $(\alpha_1^{(i)},\alpha_\tau^{(i)})=(1/2,1/2)$ (mod 1)

Finally, we analyze the case with $(\alpha^{(i)}_1,\alpha^{(i)}_\tau)=(1/2,1/2)$ in which $i_{\mathrm{max}}$ takes
\begin{align}
     i_{\mathrm{max}}=\left\{\begin{array}{cc}
         \frac{N}{2}-1,\quad\mathrm{for}\quad N=\mathrm{even}\\
         \frac{N-1}{2}-1,\quad \mathrm{for}\quad N=\mathrm{odd}\\
    \end{array}\right.
    .
\end{align}
When $N=\mathrm{even}$, the 3-point coupling is not invariant under the $\hat{U}_P^{(1/2)}$ transformation as in 
the previous cases.

We conclude that the 3-point coupling is only invariant under $\hat{U}^{(1/2)}_P$ transformation when $N=\mathrm{even}$ and trivial SS phases $(\alpha_1^{(i)},\alpha_\tau^{(i)})=(0,0)$ (mod 1).

\end{enumerate}

\bibliography{references}{}
\bibliographystyle{JHEP}

\end{document}